%% file: main.tex
\documentclass[12pt]{article}
\usepackage[utf8]{inputenc}
\usepackage[T1]{fontenc}

\usepackage{graphicx}
\graphicspath{ {./plot/} }

\usepackage{authblk}

\usepackage{caption}

\usepackage{float}
\usepackage{subcaption}
\usepackage{amsmath, amssymb, amsthm}
\usepackage{mathtools}
\usepackage{bbm}
\usepackage{bm}
\usepackage{mathrsfs}
\usepackage{tikz}
\usepackage{pgfplots}
\usepgfplotslibrary{dateplot}
\usepackage{yhmath}
\usepackage{multirow} 

\usepackage{hyperref}
\hypersetup{hidelinks}
\hypersetup{
colorlinks=true,
linkcolor=blue,
citecolor=blue
}
\usepackage{xr}

\usepackage{enumerate, enumitem}
\usepackage{fancyhdr, graphicx, proof, comment, multicol}
\usepackage[none]{hyphenat} 
\usepackage{microtype} 
\usepackage{mdframed} 

\usepackage[round]{natbib}
\setcitestyle{authoryear,open={(},close={)}}

\newtheorem{proposition}{Proposition}

\begin{document}
\def\spacingset#1{\renewcommand{\baselinestretch}%
{#1}\small\normalsize} \spacingset{1}

\newcommand{\tY}{\tilde{\mathbf{Y}}}


\title{\bf Bayesian Nonparametric Risk Assessment in Developmental Toxicity Studies with Ordinal Responses}
\author{Jizhou Kang and Athanasios Kottas\thanks{Jizhou Kang (jkang37@ucsc.edu)
is Ph.D. student, and Athanasios Kottas (thanos@soe.ucsc.edu) is Professor, Department 
of Statistics, University of California, Santa Cruz.
} \\
Department of Statistics, University of California, Santa Cruz\\
}
%
\maketitle

\bigskip
\begin{abstract}
We develop a nonparametric Bayesian modeling framework for clustered ordinal responses in 
developmental toxicity studies, which typically exhibit extensive heterogeneity. The primary 
focus of these studies is to examine the dose-response relationship, which is depicted by 
the (conditional) probability of an endpoint across the dose (toxin) levels. Standard 
parametric approaches, limited in terms of the response distribution and/or the 
dose-response relationship, hinder reliable uncertainty quantification in this context. 
We propose nonparametric mixture models that are built from dose-dependent stick-breaking 
process priors, leveraging the continuation-ratio logits representation of the multinomial 
distribution to formulate the mixture kernel. We further elaborate the modeling approach,
amplifying the mixture models with an overdispersed kernel which offers enhanced 
control of variability. We conduct a simulation study to demonstrate the benefits 
of both the discrete nonparametric mixing structure and the overdispersed kernel in 
delivering coherent uncertainty quantification. Further illustration is provided 
with different forms of risk assessment, using data from a toxicity experiment on 
the effects of ethylene glycol.
\end{abstract}

\noindent%
{\it Keywords:} Bayesian nonparametrics; Developmental toxicology; 
Dose-response relationship; Logit stick-breaking prior; Overdispersion. 

\newpage
\spacingset{1.75}

\input{chapter1}

\input{chapter2}

\input{chapter3}

\input{chapter4}
\input{chapter5}
\input{chapter6}

\section*{Acknowledgments}

The research was supported in part by the National Science Foundation under award SES 1950902.

\section*{Supporting Information}

The Supporting information includes: (i) MCMC algorithm details; (ii) proofs of the propositions; (iii) additional results for the data examples.  




\bibliographystyle{jasa3}
\bibliography{sample}

\input{Supplementary}

\end{document}

%% file: chapter1.tex
\section{Introduction}
\label{sec:overdintro}

\subsection{Background and Data}
\label{subsec:bgdata}

Ordinal regression with responses being a sum of ordinal variables is a common occurrence 
in biomedical studies. In such a problem, a multivariate ordinal response $\mathbf{Y}=(Y_1,\ldots,Y_C)$ is recorded, along with a covariate $\mathbf{x}$. Here, each component of $\mathbf{Y}$ is an integer between 0 and $m$, and $\sum_{j=1}^CY_j=m$. It is typically assumed that $\mathbf{Y}\sim \text{Mult}(m,\pi_1,\ldots,\pi_C)$.  We can equivalently view $\mathbf{Y}$ as the sum of $m$ ordinal variables, denoted as $\{\tilde{\mathbf{Y}}_q:q=1,\ldots,m\}$, where $\tY_q$ represents a standard univariate ordinal response, encoded by binary variables. Contrasting with $\tY$, we refer to variable of the type $\mathbf{Y}$ as the ``extended'' ordinal response. In this article, we will develop a modeling approach that deals with overdispersed $\mathbf{Y}$. That is, responses which we might expect to be of multinomial form, but which exhibit a variance larger than that predicted by the multinomial model.

Segment II developmental toxicology studies provide an important area of application in which data of the aforementioned structure are prevailing. In these studies, at each experimental dose level, a number of pregnant laboratory animals (dams) are exposed to the toxin after implantation. Typically, the number of fetuses on ordered categories (e.g. prenatal death, malformation, and normal) are recorded as the response. The main objective is to examine the dose-response curve, which is defined by the (conditional) probability of an endpoint across the dose levels. 
Other inferential objectives involve solving the inverse problem, where interest lies in estimation of the dose level that induces a specified extra risk comparing to the control dose. Regarding the latter, coherent uncertainty quantification of the dose-response relationships is the key for ensuring accuracy. 
We refer to, for example, \citet{Kuk2004} for a comprehensive discussion about developmental toxicity studies and the statistical issues therein. 

In a standard Segment II developmental toxicology experiment, at each experimental toxin level, $x_d$, a number, $n_d$ of pregnant laboratory animals (dams) are exposed to the toxin and the total number of implants, $m_{di}$, the number of non-viable fetuses (undeveloped embryos and/or prenatal deaths), $R_{di}$, and the number of live malformed (external, visceral or skeletal) pups, $y_{di}$, from each dam are recorded. We use $\mathbf{Y}_{di}=(R_{di},y_{di},m_{di}-R_{di}-y_{di})$ to denote the ordinal response, for the $i$-th animal at dose $x_d$. The data structure, $\{(x_d,\mathbf{Y}_{di}): d=1,\ldots,N;i=1,\ldots,n_d\}$ falls in the regression setting, with replicated ``extended'' ordinal responses at each value of the single covariate (toxin level). Hereinafter, we refer to this particular data structure as the extended setting.

As an example, we consider the data from a study where ethylene glycol (EG), an organic solvent, is evaluated for toxic effects in pregnant rats. The study involves three active toxin levels at 1.25, 2.5, and 5 g/kg, and a control group, with the respective number of dams assigned to each group being 28, 29, 27 and 28. The number of implants ranges from 1 to 18 across all dams and all dose levels, with 25th, 50th, and 75th percentiles given by 12, 14, and 15, respectively. We work with the version of the data given in Table 1 of \citet{Fung1998}. 

\begin{figure}[t!]
\centering
\includegraphics[width=14.8cm,height=3.7cm]{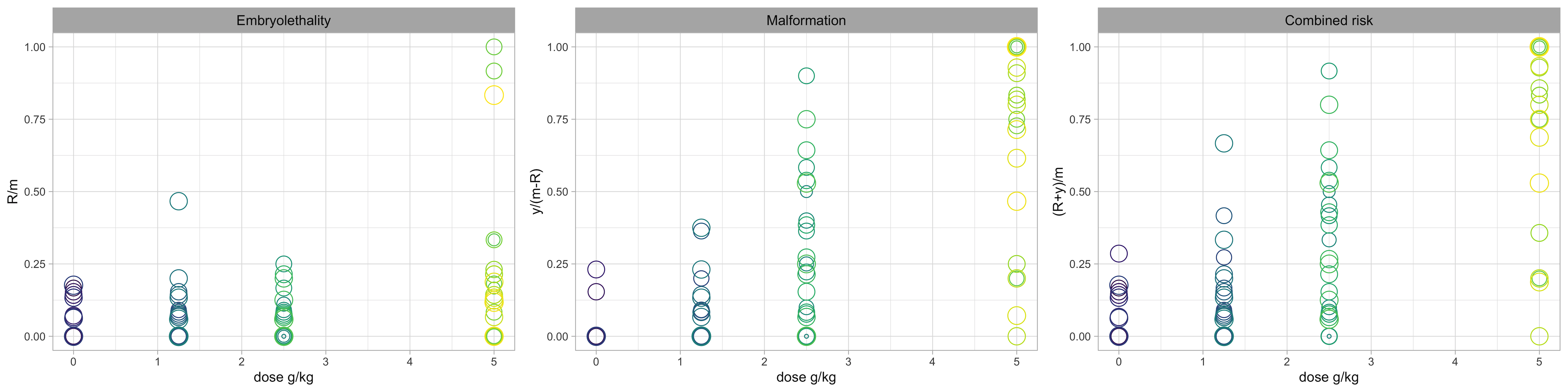}  
\caption{\small EG data. In each panel, a circle corresponds to a particular dam and the size of the circle is proportional to the number of implants. The coordinates of the circle are given by the toxin level and the proportion of the specific endpoint: non-viable fetuses among implants (left panel); malformations among live pups (middle panel); combined negative outcomes among implants (right panel).}
\label{fig:plotdata}
\end{figure}

The example data set is visualized in Figure \ref{fig:plotdata}. For each dam, we plot the observed proportions of embryolethality, malformation among live pups, and combined negative outcomes against the dose level. The color is used to facilitate identifying the same dam across panels. For the dose-response curves corresponding to these three endpoints, the empirical proportions suggest an overall increasing trend, although with no obvious parametric form for each dose-response curve. Moreover, vast variability is evident in the responses, of which the magnitude also differs across dose levels. 
Also noteworthy is a potentially different dose-response relationship for non-viable fetuses and malformed pups. A high dose usually exhibits an increase in the risk of embryolethality, while causing earlier mortality that prevents the pups surviving to be observed with malformations. Thus, it is biologically relevant to jointly model the distinct endpoints.

Because in Segment II toxicity experiments exposure occurs after implantation, we assume 
a distribution for the number of implants that does not depend on the toxin level.
Through this article, we factorize the joint distribution as 
$p(m_{di},R_{di},y_{di}\mid x_d)$ $=p(R_{di},y_{di}\mid m_{di},x_d) \, p(m_{di})$, 
and adopt a Poisson distribution for $p(m_{di})$ with support shifted such that 
$m_{di}\geq 1$. Our focus is on exploring modeling approaches for the toxin-dependent 
conditional distribution for the number of non-viable fetuses and malformations, 
$(R_{di},y_{di})$, given $m_{di}$.

\subsection{Objectives and Outline}
\label{subsec:objout}

A gold mine of modeling challenges presented in the aforementioned data structure has captured attention in the statistical literature. To address the common occurrence of overdispersed responses in developmental toxicity studies, typically used approaches involve mixture models. We investigate a spectrum of mixture models, with mixing kernel that presumes a factorized multinomial structure. Starting with continuous mixtures, we examine two popular choices, namely models based on 
the Beta-Binomial (BB) distribution and the Logistic-Normal-Binomial (LNB) distribution. 
We argue that these models preclude reliable risk assessment in this application, because of their parametric form in both the response distribution and the dose-response relationship.

Turning to discrete mixture models, those based on the Dirichlet 
process \citep[DP,][]{Ferguson1973} serve as an initial reference. In our context, since the covariate (toxin level) is involved, a more relevant modeling approach is to consider mixture model with a dependent Dirichlet process \citep[DDP,][]{MacEachern2000} or dependent stick-breaking process \citep{Quintana2022} prior.
Specifically, the nonparametric mixture models proposed 
in \citet{KangKottas2022} 
provide the foundation for our modeling approach.
Here, enhanced flexibility is achieved through a nonparametric mixture 
of continuation-ratio logits factorization of multinomial distributions \citep{Tutz1991}, mixing through 
a dependent stick-breaking process prior placed on the probability parameters. Nonetheless, 
its kernel is restricted in terms of modeling extended ordinal response, providing opportunities 
for novel models which enable more effective control of the ordinal responses' variability.

We consider a combination of these two types of mixture models. Specifically, we adopt a continuous mixture model as the kernel, which is then encapsulated in the discrete nonparametric mixing structure. The derived models inherit flexibility from the discrete mixture models, while potentially allowing an improvement in accounting for 
overdispersion through the extra set of parameters introduced in the kernel. Regarding the choice of kernel, the LNB distribution is preferred from a computational efficiency consideration. Besides expanding a developed approach, motivation for examining the new model also originates from regulatory guidance \citep{EPA1991}, which requires considering an adequate set of models for developmental toxicity risk assessment.      

Traditional approaches are based on variations of the trinomial distribution that 
enable characterizing overdispersion to describe the joint conditional probability 
$p(R_{di},y_{di}\mid m_{di},x_d)$ \citep{Chen1991,Krewski1994}. An alternative approach involves employing the Rao-Scott transformation, which eliminates overdispersion in the transformed data, enabling the application of the standard trinomial model \citep{Krewski1995}. Seeking more flexibility in response distributions, models have been developed from a frequentist perspective \citep{Pang2005,Wu2014} and a Bayesian perspective \citep{Dunson2003}, focusing primarily on binary outcomes. Nonetheless, such models still make strong assumptions about the dose-response 
relationships. 

Bayesian nonparametric methods have been explored as a powerful tool for analysis of development toxicology data. Focusing on studies that involve a discrete response, \citet{Dominici2001} proposed a product of Dirichlet process (DP) mxitures approach to deal with combined negative outcomes. Targeting the same type of responses, \citet{KassieKottas2014} built a nonparametric mixture model from a dependent Dirichlet process (DDP) prior, with the dependence of the mixing distributions governed by the dose level. Models that jointly consider various types of responses have also been explored, including, for binary and continuous responses \citep{Hwang2014}, categorical and continuous responses \citep{KassieKottas2017}, binary and continuous responses and litter size \citep{Hwang2018}. The most relevant methodology is the one discussed in \citet{KF2013}, which deals with ordinal responses. 
They use a product of Binomials as the kernel, to capture the nested structure of the responses, and a common-weights DDP prior for the dose-dependent mixing distributions. We develop different (including more general) mixture models than the one in \citet{KF2013}.

The rest of this article is organized as follows. In Section \ref{sec:contmixmodel}, we review the two typical continuous mixture models for accounting for overdispersion in ordinal responses, and demonstrate their limitations in uncertainty quantification with the EG data. The discrete nonparametric mixture models with either type of kernel are formulated and examined in depth in Section \ref{sec:discmixmodel}. Section \ref{sec:simstudy}
introduces two carefully designed simulation studies that reflect our main contributions. We compare the performance of the nonparametric mixture models through a series of risk assessments, conducted on the EG data. The main results are presented in Section \ref{sec:dtdataill}. Finally, Section \ref{sec:sumrmks} concludes with a summary and discussion.

%% file: chapter2.tex
\section{Continuous Mixture Models}
\label{sec:contmixmodel}

This section focuses on providing an appropriate context for the models that will be examined later in this article. We start by reviewing properties of the classic Beta-Binomial and Logistic-Normal-Binomial distribution. Models built on them for ordinal responses in developmental toxicity study are also discussed.

\subsection{Beta-Binomial and Logistic-Normal-Binomial}

Both the BB and the LNB distribution can be viewed as a continuous mixture of the Binomial distribution. Consider modeling the number of positive responses, denoted by $Y$, among $m$ trials. Then, the BB model assumes
\begin{equation*}
	Y\mid m,\theta,\lambda \,\sim\, BB(m,\theta,\lambda) \, = \,
\int Bin(Y\mid m,\psi)Beta(\psi\mid \lambda\varphi(\theta),\lambda(1-\varphi(\theta))) \, d\psi.
\end{equation*}
On the other hand, the LNB model is formulated as
\begin{equation*}
	Y\mid m,\theta,\sigma^2 \,\sim\, LNB(m,\theta,\sigma^2) \, = \,
 \int Bin(Y\mid m,\varphi(\psi))N(\psi\mid \theta,\sigma^2) \, d\psi.
\end{equation*}
Here, $\varphi(x) = $ $\exp(x)/(1 + \exp(x))$ denotes the standard logistic function.
Under a regression setting, covariate effects can be incorporated into the model by setting $\theta=\theta(\mathbf{x})$.

For a deeper comprehension of these distributions, we consider the alternative encoding of $Y$ with binary indicators $\{\tilde{Y}_q:q=1,\ldots,m\}$, such that $Y=\sum_{q=1}^m\tilde{Y}_q$. Both the BB model and the LNB model postulate exchangeability, in lieu of independence, for $\tilde{Y}_q$, which induces marginal dependence among them. Capitalizing on overdispersion results for mixtures from exponential families \citep{Shaked1980}, we can show that the variance of $Y$ under either of the models is larger than the variance of $Y$ under a Binomial model, that is, the mixture models achieve overdispersion. The extent of overdispersion is controlled by the correlation between any pair of $\tilde{Y}_q$ and $\tilde{Y}_{q^{\prime}}$, for $q,q^{\prime}\in\{1,\ldots,m\}$. 

Under the BB distribution, $\text{E}(\tilde{Y}_q\mid \theta)$ $=\varphi(\theta)$, which is the same as the mean under a Binomial distribution. For the correlation, $\text{Corr}(\tilde{Y}_q,\tilde{Y}_{q^{\prime}}\mid \lambda)=(1+\lambda)^{-1}$. Therefore, $\lambda$ controls the dependence among $\tilde{Y}_q$, hence the variance of $Y$, and is termed the overdispersion parameter. 

Because the logit-normal integral in general does not have analytical form, neither $\text{E}(\tilde{Y}_q\mid m,\theta,\sigma^2)$ nor $\text{Corr}(\tilde{Y}_q,\tilde{Y}_{q^{\prime}}\mid \theta,\sigma^2)$ are available in closed form for the LNB distribution. Nonetheless, we have the following approximation based on a second-order Taylor series expansion, which helps conceptualize the distribution.

\begin{proposition}
\label{prop:lnbmargapprox}
	Suppose $\tilde{Y}_q\mid\psi\stackrel{i.i.d.}{\sim}Bern(\varphi(\psi))$, for $q=1,\ldots,m$, and $\psi\mid\theta,\sigma^2\sim N(\theta,\sigma^2)$. Then, marginalizing over $\psi$, 
	\begin{equation}
	\begin{array}{rcl}
			 \text{E}(\tilde{Y}_q\mid \theta,\sigma^2) & \approx & \varphi(\theta)+\frac{\sigma^2}{2}\varphi^{\prime\prime}(\theta)\\
			 \text{Corr}(\tilde{Y}_q,\tilde{Y}_{q^{\prime}}\mid \theta,\sigma^2) & \approx &
			 \frac{\sigma^2\varphi^{\prime}(\theta)[4-\sigma^2(1-2\varphi(\theta))^2]}{4+\sigma^2(1-2\varphi(\theta))[2-4\varphi(\theta)-\sigma^2\varphi^{\prime\prime}(\theta)]}.
	\end{array}
	\label{eq:approxlnb}
	\end{equation}
\end{proposition}

The proof is shown in the Supporting Information. Proposition \ref{prop:lnbmargapprox} reveals features of the LNB distribution, contrasting to the BB distribution, in two folds. Firstly, the LNB model introduces a fluctuation in the mean, with the magnitude managed by $\sigma^2$. Besides, both $\sigma^2$ and $\theta$ affect the correlation. To be aligned with the BB distribution, we term $\sigma^2$ the overdispersion parameter. Note however that the overdispersion parameter $\lambda$ of the BB distribution affects only the variance of $Y$, while the LNB distribution $\sigma^2$ parameter influences both the mean and variance of $Y$.

In terms of Bayesian inference, the LNB model is more attractive. When the priors for hyperparameters are conditionally conjugate, leveraging the Pólya-Gamma data augmentation approach \citep{Polson2013}, we can obtain posterior samples of parameters through Gibbs sampling. 
On the contrary, Bayesian implementation of the BB model requires tuning Metropolis-Hasting samplers. For this reason, we choose the LNB distribution as the building block for the nonparametric mixture model with overdispersed kernel discussed in Section \ref{subsec:bnpoverdmodel}.

\subsection{Models for Extended Ordinal Responses}

In developmental toxicity studies, the developing fetuses are at risk of fetal death due to the toxin insult. For those who survive the entire gestation period, malformation may be exhibited. The sequential nature of the response suggests factorizing the joint distribution as $p(R,y\mid m)=p(R\mid m)p(y\mid R,m)$. Let $\mathbf{x}=(1,x)$, where $x$ denotes the dose level. We omitted the subscript $d$ and $i$ for notation simplicity. The model that assumes each part of the factorization following a BB distribution is given by
\begin{equation}
	(R,y)\mid m,\theta_1(\mathbf{x}),\theta_2(\mathbf{x}),\boldsymbol{\lambda}\,\sim\, BB(R\mid m,\theta_1(\mathbf{x}),\lambda_1)BB(y\mid m-R,\theta_2(\mathbf{x}),\lambda_2),
	\label{eq:paramcrbb}
\end{equation}
where $\theta_j(\mathbf{x})=\mathbf{x}^{\top}\boldsymbol{\beta}_j$, $j=1,2$, and $\boldsymbol{\lambda}=(\lambda_1,\lambda_2)$. Here, we assume a separate mixing distribution for each component, which is the key for effective interpretation and implementation of the model. Because of its induced interpretation for the response probabilities, the factorization is termed continuation-ratio in the literature. Accordingly, we term (\ref{eq:paramcrbb}) the continuation-ratio Beta-Binomial (``CR-BB'') model.  Similarly, if the LNB distribution is used for each part of the factorization, the model is formulated as
\begin{equation}
	(R,y)\mid m,\theta_1(\mathbf{x}),\theta_2(\mathbf{x}),\boldsymbol{\sigma^2}\,\sim\, LNB(R\mid m,\theta_1(\mathbf{x}),\sigma_1^2)LNB(y\mid m-R,\theta_2(\mathbf{x}),\sigma_2^2),
	\label{eq:paramcrlnb}
\end{equation}
where $\boldsymbol{\sigma^2}=(\sigma_1^2,\sigma_2^2)$, and it will be referred to as the continuation-ratio Logistic-Normal-Binomial (``CR-LNB'') model.

To aid in exploring the relationship between the toxin level and the probability of the various endpoints, it is helpful to consider the underlying binary responses. In particular,  
for a generic dam with $m$ implants exposed 
to toxin level $x$, we denote by $\boldsymbol{\tilde{R}}=$ $\{\tilde{R}_q:q=1,\cdots,m\}$ the non-viable fetus indicators,
and $\boldsymbol{\tilde{y}}=$ $\{\tilde{y}_{l}: l=1,\cdots,m-\sum_{q=1}^m\tilde{R}_q\}$ the malformation indicators
for the live pups, such that the extended ordinal response is $\mathbf{Y}=(R,y,m-R-y)$,
where $R = \sum_{q=1}^m\tilde{R}_q$ and $y = \sum_{l=1}^{m-R} \tilde{y}_{l}$. The dose response curves are defined with the alternative encoding of the responses. Following the standard risk assessment methods in the literature \citep[e.g.][]{Krewski1995}, we consider the dose-response curves of embryolethality, malformation of viable fetus, and combined risk, implicitly conditioning on $m=1$ and the model $\mathcal{M}$, defined respectively as
$D(x) =$ $\text{Pr}(\tilde{R}=1\mid x)$,
$M(x) =$ $\text{Pr}(\tilde{y}=1\mid \tilde{R}=0,x)$,
and $r(x) =$ $\text{Pr}(\tilde{R}=1 \,\, \text{or} \,\, \tilde{y}=1\mid x) =$ 
$\text{Pr}(\tilde{R}=0 \,\, \text{and} \,\, \tilde{y}=1\mid x)$ + 
$\text{Pr}(\tilde{R}=1\mid x)$. 

\begin{figure}[t!]
\centering
\begin{subfigure}{0.485\textwidth}
  \centering 
  \includegraphics[width=7.3cm,height=2.43cm]{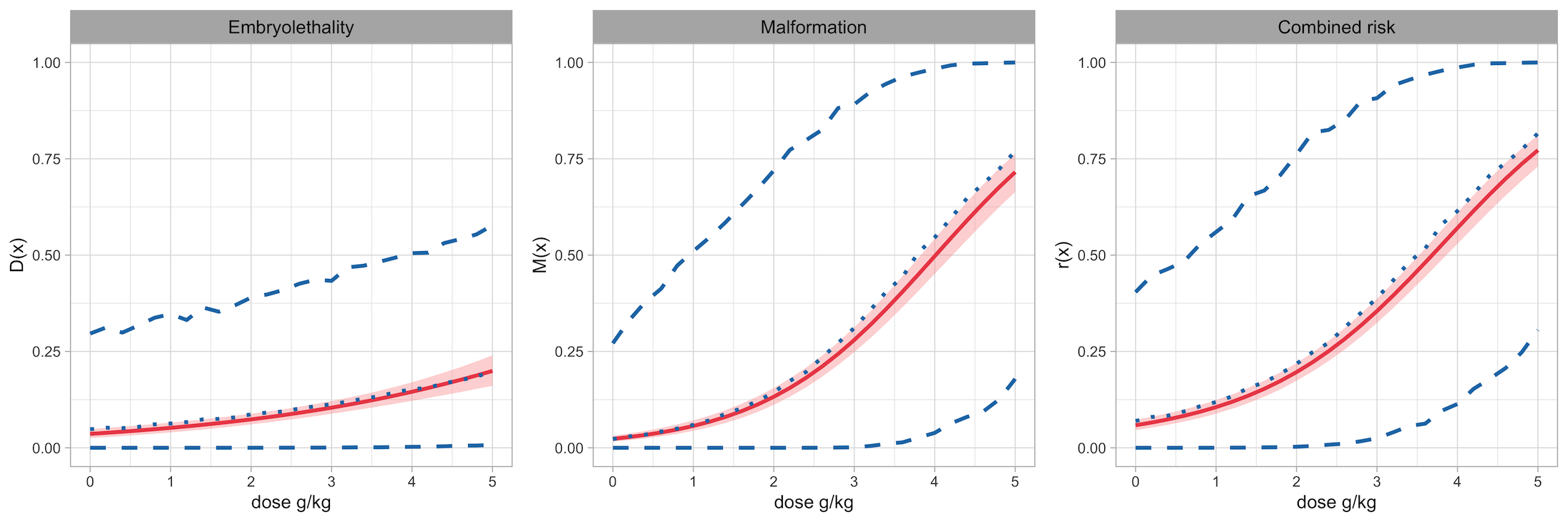}  
  \caption{\small ``CR-BB'' model.}
  \label{subfig:egbb}
\end{subfigure}
\hfill
\begin{subfigure}{0.485\textwidth}
  \centering
  \includegraphics[width=7.3cm,height=2.43cm]{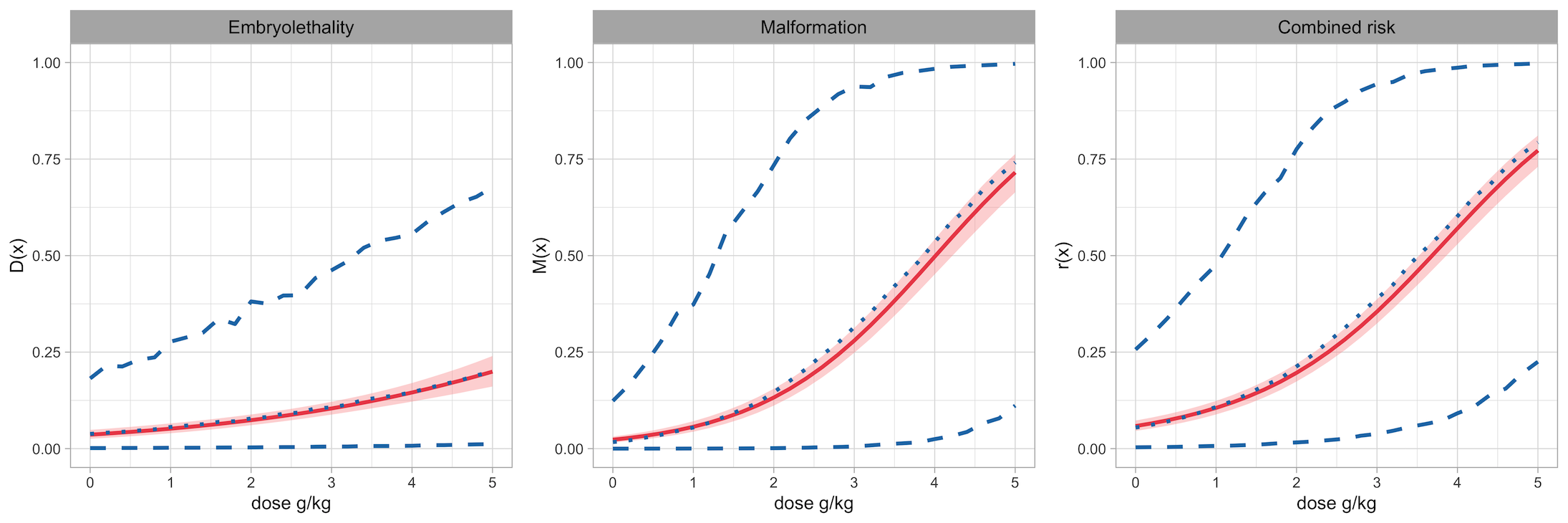}  
  \caption{\small ``CR-LNB'' model.}
  \label{subfig:eglnb}
\end{subfigure}
\caption{\small EG data. Posterior mean (dotted line) and 95\% interval estimate (dashed lines) for the dose response curves. The red solid line and shaded region is the posterior mean and 95\% interval estimates obtained under a continuation-ratio logits model. }
\label{fig:egparam}
\end{figure}

For the EG data, we fit the ``CR-BB'' model and the ``CR-LNB'' model to obtain posterior inference for the dose-response curves. The resulting point and interval estimates are displayed in Figure \ref{fig:egparam}, where, as a reference point, we also present the same inference under the continuation-ratio logits model. Without a mixing structure, the continuation-ratio logits model cannot account for overdispersion, leading to very narrow uncertainty bands. In contrast, the ``CR-BB'' and ``CR-LNB'' models provide overly wide interval estimates. This pattern emerges because the continuous mixture models pool the variability over the dose range, providing significant uncertainty even at dose levels with relatively small observed heterogeneity. Moreover, due to their parametric form, these models tend to overcompensate for the data heterogeneity by increasing the variability in the response distribution.

Such limitations of parametric continuous mixture models motivate us to consider discrete nonparametric mixture models, specifically the mixing structure induced by a dose-dependent stick-breaking process prior. By permitting clustered mixing parameters, the discrete mixture models have the potential to manage the variability of response distribution more effectively. Next, we explore modeling approaches in this direction.

%% file: chapter3.tex
\section{Discrete Mixture Models}
\label{sec:discmixmodel}

\subsection{Models with Continuation-ratio Logits Kernel}  
\label{subsec:bnpcrbin}



We consider a generalization of the continuation-ratio logits regression model via Bayesian nonparametric mixing. 
The model extension is achieved through a covariate-dependent nonparametric prior, $G_{\mathbf{x}}=$
$\sum_{\ell=1}^{\infty} \omega_{\ell}(\mathbf{x}) \, \delta_{(\theta_{1\ell}(\mathbf{x}),\theta_{2\ell}(\mathbf{x}))}$, leading to the general model
\begin{equation}
	(R,y)\mid m,G_{\mathbf{x}} \, \sim \, \sum_{\ell=1}^{\infty} \omega_{\ell}(\mathbf{x}) \, 
Bin(R \mid m,\varphi({\theta}_{1\ell}(\mathbf{x})))Bin(y\mid m-R, \varphi({\theta}_{2\ell}(\mathbf{x}))).
\label{eq:dtsmodelform}
\end{equation}

As discussed in \citet{KangKottas2022}, the logit stick-breaking process (LSBP) prior \citep{RigonDurante2021} has a structural similarity with the continuation-ratio logits, which offers key advantages in model properties and implementation. 
We assume the following LSBP prior for the covariate-dependent weights:
\begin{equation}
\omega_{1}(\mathbf{x}) \, = \, \varphi(\mathbf{x}^{\top}\boldsymbol{\gamma}_1), \,\,
\omega_{\ell}(\mathbf{x}) \, = \,
\varphi(\mathbf{x}^{\top}\boldsymbol{\gamma}_{\ell}) 
\prod_{h=1}^{\ell-1} (1-\varphi(\mathbf{x}^{\top}\boldsymbol{\gamma}_h)), \,\, \ell \geq 2; 
\,\,
\boldsymbol{\gamma}_{\ell} \, \stackrel{i.i.d.}{\sim} \, N(\boldsymbol{\gamma}_0,\Gamma_0)
\label{eq:dtsgenweights}
\end{equation}
In addition, the atoms are built through a linear regression structure, 
\begin{equation}
\theta_{j\ell}(\mathbf{x}) \, = \, \mathbf{x}^{\top}\boldsymbol{\beta}_{j\ell}
\mid \boldsymbol{\mu}_j, \Sigma_j \, \stackrel{ind.}{\sim} \, 
N(\mathbf{x}^{\top}\boldsymbol{\mu}_j,\mathbf{x}^{\top}\Sigma_j\mathbf{x}), \,\,\,\,\,
j=1,2, \,\,\, \ell \geq 1,
\label{eq:dtsgenatoms}
\end{equation}
with the random variables that define the atoms assumed a priori independent of those 
that define the weights. The model is completed with the conjugate prior for the collection of hyperparameters 
$\boldsymbol{\psi} =$ $\{\boldsymbol{\mu}_j,\Sigma_j: j=1,2\}$, that is, 
\begin{equation}
\Sigma_j \stackrel{ind.}{\sim} IW(\nu_{0j},\Lambda_{0j}^{-1}), \,\,\,\,\,\,\,
\boldsymbol{\mu}_j \mid \Sigma_j \stackrel{ind.}{\sim} 
N(\boldsymbol{\mu}_{0j},\Sigma_j/\kappa_{0j}), \,\,\,\,\, j=1,2.
\label{eq:dtsgenhyperparameters}
\end{equation}
We refer to the discrete mixture model in (\ref{eq:dtsmodelform}), with mixing weights and atoms specified respectively in (\ref{eq:dtsgenweights}) and (\ref{eq:dtsgenatoms}), as the general mixture of product of Binomials kernel (``Gen-Bin'') model.

We establish a useful connection of the nonparametric mixture model built for extended ordinal response, with a model built for the underlying response $\boldsymbol{\tilde{R}}$ and $\boldsymbol{\tilde{y}}$. Using the same nonparametric prior specified in (\ref{eq:dtsgenweights}) and (\ref{eq:dtsgenatoms}), together with a product of Bernoullis kernel, the nonparametric mixture model for underlying binary response can be formulated as
\begin{equation}
	(\boldsymbol{\tilde{R}},\boldsymbol{\tilde{y}})\mid m,G_{\mathbf{x}}\, \sim \,\sum_{\ell=1}^{\infty} \omega_{\ell}(\mathbf{x}) \, 
\prod_{q=1}^m Bern(\tilde{R}_q\mid \varphi({\theta}_{1\ell}(\mathbf{x})))\prod_{l=1}^{m-\sum_q\tilde{R}_q}Bern(\tilde{y}_l\mid \varphi({\theta}_{2\ell}(\mathbf{x}))).
\label{eq:dtsgenprodbern}
\end{equation}   
We can show that the mixture models (\ref{eq:dtsmodelform}) and (\ref{eq:dtsgenprodbern}) are
equivalent in the sense that the moment generating function (MGF) of $(R,y)$ under (\ref{eq:dtsmodelform}) is equal to the MGF of $(\sum \tilde{R}_q,\sum \tilde{y}_l)$ under (\ref{eq:dtsgenprodbern}). The result is formally stated in Proposition \ref{prop:equmgforimodel}, with the proof presented in the Supporting Information. 

\begin{proposition}
\label{prop:equmgforimodel}
Let $\mathcal{M}$ and $\tilde{\mathcal{M}}$ denote the mixture models (\ref{eq:dtsmodelform}) and (\ref{eq:dtsgenprodbern}), respectively. With the same $m$, and $G_{\mathbf{x}}$ formulated by (\ref{eq:dtsgenweights}) and (\ref{eq:dtsgenatoms}), 
\begin{equation}
	E_{\mathcal{M}}(e^{t_1R+t_2y}\mid m,G_{\mathbf{x}})\,=\, E_{\tilde{\mathcal{M}}}(e^{t_1\sum \tilde{R}_q+t_2\sum \tilde{y}_l}\mid m,G_{\mathbf{x}})
	\label{eq:equalMGF}
\end{equation}
The subscript of the expectation refers to the distribution under which the expectation is taken. 	
\end{proposition}


Proposition \ref{prop:equmgforimodel} allows us to examine the dose-response curves for a dam with a generic number of fetuses, which of course includes $m=1$. Consequently, the expressions for the dose-response curves of embryolethality $D(x)$, malformation $M(x)$, and combined risk $r(x)$, under the proposed model, are given by
\begin{equation}
	\begin{split}
		D(x)&\, = \, \text{Pr}(\tilde{R}=1\mid G_{\mathbf{x}})\,=\,\sum_{\ell=1}^{\infty} 
\omega_{\ell}(\mathbf{x}) \, \varphi(\theta_{1\ell}(\mathbf{x}));\\
    M(x)&\, = \, \text{Pr}(\tilde{y}=1\mid\tilde{R}=0,G_{\mathbf{x}})\,=\,
     \sum_{\ell=1}^{\infty} \frac{\omega_{\ell}(\mathbf{x})
[1-\varphi(\theta_{1\ell}(\mathbf{x}))]}
{\sum_{\ell=1}^{\infty}\omega_{\ell}(\mathbf{x})
[1-\varphi(\theta_{1\ell}(\mathbf{x}))]} \, 
\varphi(\theta_{2\ell}(\mathbf{x}));\\
    r(x)&\, = \, \text{Pr}(\tilde{R}=1 \,\, \text{or} \,\, \tilde{y}=1\mid G_{\mathbf{x}})\,=\,
    1-\sum_{\ell=1}^{\infty} 
\omega_{\ell}(\mathbf{x}) \, [1-\varphi(\theta_{1\ell}(\mathbf{x}))]
[1-\varphi(\theta_{2\ell}(\mathbf{x}))],
	\end{split}
	\label{eq:dosecurvegenori}
\end{equation}
with $\omega_{\ell}(\mathbf{x})$ and $\theta_{1\ell}(\mathbf{x})$, $\theta_{2\ell}(\mathbf{x})$ defined in (\ref{eq:dtsgenweights}) and (\ref{eq:dtsgenatoms}), respectively. Note that all three dose-response curves admit a weighted sum representation with covariate-dependent weights, which enables local adjustment over the dose level, resulting in flexible estimation of the dose-response relationships.

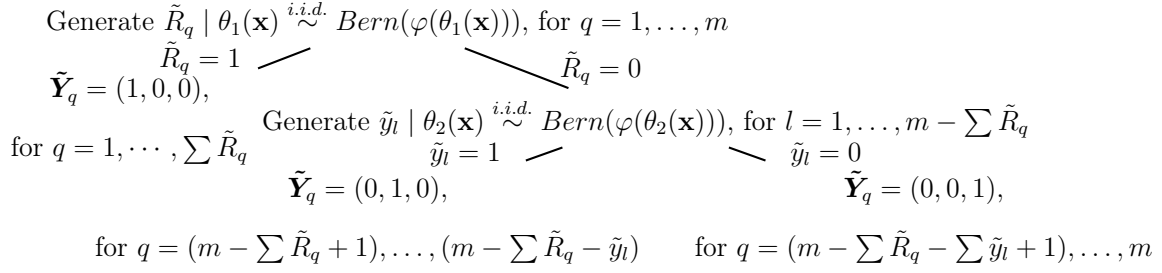
\begin{figure}[t!]
\centering
\begin{tikzpicture}[thick,scale=0.88, every node/.style={scale=0.88}]

\node {Generate $\tilde{R}_q\mid \theta_{1}(\mathbf{x})\stackrel{i.i.d.}{\sim} Bern(\varphi(\theta_{1}(\mathbf{x})))$, for $q=1,\ldots,m$}[sibling distance = 7.8cm]
    child {node [align=center] {$\boldsymbol{\tilde{Y}}_q=(1,0,0)$, \\for $q=1,\cdots,\sum\tilde{R}_q$} edge from parent node [left,xshift=-0.5cm] {$\tilde{R}_q=1$}}
    child {node {Generate $\tilde{y}_l\mid \theta_{2}(\mathbf{x})\stackrel{i.i.d.}{\sim} Bern(\varphi(\theta_2(\mathbf{x})))$, for $l=1,\ldots,m-\sum\tilde{R}_q$}[sibling distance = 8.4cm] 
    child {node [align=center] {$\boldsymbol{\tilde{Y}}_q=(0,1,0)$,\\for $q=(m-\sum\tilde{R}_q+1),\ldots,(m-\sum\tilde{R}_q-\tilde{y}_l)$} edge from parent node [left,xshift=-0.5cm] {$\tilde{y}_l=1$}}
    child {node [align=center] {$\boldsymbol{\tilde{Y}}_q=(0,0,1)$, \\for $q=(m-\sum\tilde{R}_q-\sum\tilde{y}_l+1),\ldots,m$}
    edge from parent node [right,xshift=0.5cm] {$\tilde{y}_{l}=0$}}
    edge from parent node [right,xshift=0.5cm] {$\tilde{R}_{q}=0$}};
\end{tikzpicture}
\caption{
\small Connection between alternative encodings of the ordinal response.
}
\label{fig:dtsseqbreak}
\end{figure}

Another equivalent encoding of the responses comes from the connection between the standard and extended ordinal response. 
Indeed, let $\{\boldsymbol{\tilde{Y}}_q:q=1,\cdots,m\}$ be a collection of standard ordinal responses. 
That is, $\boldsymbol{\tilde{Y}}_q=(\tilde{Y}_{q1},\tilde{Y}_{q2},\tilde{Y}_{q3})$, where $\tilde{Y}_{qj}$ are binary, and only one $\tilde{Y}_{qj}=1$, for $j=1,2,3$. 
We can view $\boldsymbol{\tilde{Y}}_q$ as the ordinal response from an implant of the dam. 
They are linked with $\mathbf{Y}=(R,y,m-R-y)$ through $\mathbf{Y}=\sum_{q=1}^m\boldsymbol{\tilde{Y}}_q$.
In addition, $\boldsymbol{\tilde{Y}}_q$ are connected with $\tilde{R}_q$ and $\tilde{y}_l$ through the sequential mechanism of the continuation-ratio logits structure, depicted in Figure \ref{fig:dtsseqbreak}. Introducing $\boldsymbol{\tilde{Y}}_q$ facilitates the study of overdispersion.

In the context of development toxicology studies, responses from the fetuses within the same dam are typically assumed to be positively correlated, resulting in overdispersion. Therefore, relevant modeling methods should promote positive intracluster correlations. Here, the cluster refers to the dam. Under the proposed model, the intracluster correlation at category $j$, $j=1,2,3$, for any implants $q$ and $q^{\prime}$ from the same dam, is given by
\begin{equation}
\text{Corr}(\tilde{Y}_{qj},\tilde{Y}_{q^{\prime}j}\mid G_{\mathbf{x}})=\frac{\text{E}(\tilde{Y}_{qj}\tilde{Y}_{q^{\prime}j}\mid G_{\mathbf{x}})-\text{E}(\tilde{Y}_{qj}\mid G_{\mathbf{x}})
\text{E}(\tilde{Y}_{q^{\prime}j}\mid G_{\mathbf{x}})}
{\{ \text{Var}(\tilde{Y}_{qj}\mid G_{\mathbf{x}})\text{Var}(\tilde{Y}_{q^{\prime}j}\mid G_{\mathbf{x}}) \}^{1/2}},
\label{eq:intracorrorimodel}
\end{equation}
where $\text{E}(\tilde{Y}_{qj} \mid G_{\mathbf{x}})=$ 
$\text{E}(\tilde{Y}_{q^{\prime}j}\mid G_{\mathbf{x}})=$
$\sum_{\ell=1}^{\infty}\omega_{\ell}(\mathbf{x})\,\{\varphi(\theta_{j\ell}(\mathbf{x}))
\prod_{k=1}^{j-1}[1-\varphi(\theta_{k\ell}(\mathbf{x}))]\}$, 
$\text{Var}(\tilde{Y}_{qj} \mid G_{\mathbf{x}})=$
$\text{Var}(\tilde{Y}_{q^{\prime}j}\mid G_{\mathbf{x}})=$
$\text{E}(\tilde{Y}_{qj}\mid G_{\mathbf{x}})-[\text{E}(\tilde{Y}_{qj}\mid G_{\mathbf{x}})]^2$, 
and $\text{E}(\tilde{Y}_{qj}\tilde{Y}_{q^{\prime}j}\mid G_{\mathbf{x}})=$
$\sum_{\ell=1}^{\infty}\omega_{\ell}(\mathbf{x})$          $\{\varphi(\theta_{j\ell}(\mathbf{x}))\, \prod_{k=1}^{j-1}[1-\varphi(\theta_{k\ell}(\mathbf{x})]\}^2$, with $\varphi(\theta_{3\ell}(\mathbf{x}))\equiv 1$. \cite{KassieKottas2014} have shown that the intracluster correlation is positive under a 
common-weights DDP mixture of Binomial distributions. The required assumptions are that 
the variance, $\text{Var}(\tilde{Y}_{qj}\mid G_{\mathbf{x}})$, and correlation, $\text{Corr}(\tilde{Y}_{qj},\tilde{Y}_{q^{\prime}j}\mid G_{\mathbf{x}})$, are common 
within the cluster. These assumptions hold here, since any pair of $\tilde{Y}_{qj}$, $\tilde{Y}_{q^{\prime}j}$ are associated with the same dose level. As a result, the positive intracluster correlations result extends to our case, i.e., $\text{Corr}(\tilde{Y}_{qj},\tilde{Y}_{q^{\prime}j}\mid G_{\mathbf{x}})>0$, $\forall\, j$.

A practically relevant modeling aspect revolves around possible monotonicity restrictions 
for the dose-response functions. Developmental toxicity studies involve a small number of
administered toxin levels. Hence, under nonparametric mixture models for the categorical 
responses, a monotonic trend in the prior expectation for the dose-response curves is desirable for effective interpolation and extrapolation inference. This is discussed in \citet{KF2013}
and \cite{KassieKottas2014} under common-weights DDP mixture models, and is also relevant 
in our model setting. 
Using the prior specification strategy proposed in \citet{KangKottas2022}, 
we can incorporate a non-decreasing trend in the prior expected dose-response curves.
We note however 
that prior (and thus posterior) realizations for the dose-response curves are not structurally 
restricted to be non-decreasing.

Two simplifications of the general model are discussed in \citet{KangKottas2022}, namely the common-weights model and the common-atoms model. Due to the monotonicity restriction of the prior expectation for the dose-response curves, the common-atoms model  is not a practical option. This is because the common-atoms model adjusts the shape of dose-response curves only through the weights, resulting in prior expectations that are constant with respect to the toxin level covariate. 
 Nonetheless, the common-weights model is worth exploring, because it bridges the general nonparametric mixture model proposed here with the model discussed in \citet{KF2013}. The common-weights mixture with product of Binomial kernels (``CW-Bin'') model is specified as
\begin{equation*}
	(R,y)\mid m,G_{\mathbf{x}} \, \sim \, \sum_{\ell=1}^{\infty} \omega_{\ell} \, 
Bin(R \mid m,\varphi({\theta}_{1\ell}(\mathbf{x})))Bin(y\mid m-R, \varphi({\theta}_{2\ell}(\mathbf{x}))),
\label{eq:dtscommonweights}
\end{equation*}
with $\omega_{1} = V_{1}$, and $\omega_{\ell} =$ 
$V_{\ell} \prod_{h=1}^{\ell-1} (1 - V_{h})$, for $\ell \geq 2$, where 
$V_{\ell} \mid \alpha \stackrel{i.i.d.}{\sim} \, Beta(1,\alpha)$, and ${\theta}_{1\ell}(\mathbf{x})$, ${\theta}_{2\ell}(\mathbf{x})$ defined in (\ref{eq:dtsgenatoms}). \citet{KF2013} adopt the same structure for the weights in their mixture model, while the atoms are chosen as Gaussian processes with the mean function postulating a linear regression form. They extend the common-weights model by incorporating a more flexible structure for the atoms. Note that their model still does not allow the dose-response curves for the embryolethality and combined negative outcome to have dose-dependent weights, which is an asset of our general model.  

For Markov chain Monte Carlo (MCMC) posterior simulation, we notice that
the blocked Gibbs sampler \citep{IshwaranJames2001} proposed in \citet{KangKottas2022} is also applicable to conduct posterior simulation with the ``Gen-Bin'' model and the ``CW-Bin'' model. 
Posterior realizations for the dose-response curves and intracluster correlations can be obtained by evaluating the corresponding expressions with MCMC posterior samples of model parameters. Moreover, for each endpoint, we can obtain the posterior distribution of a calibrated dose level for a specified probability, by (numerically) inverting the posterior realization of the corresponding dose-response curve. We illustrate the procedure with the EG data in Section \ref{sec:dtdataill}.

The discrete mixing structure in conjunction with the restricted kernel implies \textit{a priori} a trade-off between the variability of the response and the variability of the dose-response curve. 
Because overdispersion is not admitted in the kernel, the mixture model seeks to account for the vast variability in the response by activating more effective components. Contrarily, because of the discrete mixture structure, more effective components lead to less variability in the prior realizations of dose response curves, yielding overconfident prior intervals.
Seeking coherent uncertainty quantification for both the response distribution and the dose-response curves, we consider building discrete mixture models with a kernel that allows higher level of dispersion.



\subsection{Models with Overdispersed Kernel}
\label{subsec:bnpoverdmodel}

Parallel to the development of the ``Gen-Bin'' model, we formulate the alternative modeling approach with overdispersed kernel starting from its parametric backbone in (\ref{eq:paramcrlnb}). Amplified with the general dose-dependent nonparametric prior 
we obtain
\begin{equation}
	(R,y)\mid m,G_{\mathbf{x}},\boldsymbol{\sigma^2} \, \sim \, \sum_{\ell=1}^{\infty} \omega_{\ell}(\mathbf{x}) \, 
LNB(R \mid m,{\theta}_{1\ell}(\mathbf{x}),\sigma^2_1)LNB(y\mid m-R, {\theta}_{2\ell}(\mathbf{x}),\sigma_2^2),
\label{eq:overdgenmodelform}
\end{equation} 
The prior on the weights $\omega_{\ell}(\mathbf{x})$ is specified as the same LSBP prior given in (\ref{eq:dtsgenweights}), while the atoms $\theta_{1\ell}(\mathbf{x})$, $\theta_{2\ell}(\mathbf{x})$ and their prior are specified as in (\ref{eq:dtsgenatoms}) and (\ref{eq:dtsgenhyperparameters}). The model formulation is completed with $\sigma_j^2\stackrel{i.i.d.}{\sim}IG(a_{\sigma},b_{\sigma})$, for $j=1,2$. This model formulation shall be referred to as the general mixture with product of LNB kernel (``Gen-LNB'') model hereinafter.

The ``Gen-LNB'' model includes both the ``CR-LNB'' model and the ``Gen-Bin'' model as special (limiting) cases. If $\boldsymbol{\gamma}_1$ is such that $\varphi(\mathbf{x}^T\boldsymbol{\gamma}_1)$ is effectively equal to one, 
the nonparametric model collapses to the model in (\ref{eq:paramcrlnb}). If we let $\sigma_j^2\to 0^+$ for $j=1,2$, the kernel collapses to the continuation-ratio logits model, resulting in the ``Gen-Bin'' model. The specific mixing structure allows smooth deviations from the Binomial, while keeping the extra level of flexibility, brought in by the discrete (infinite) mixture.

To investigate model properties, we build its connection with the nonparametric mixture model for the underlying $\boldsymbol{\tilde{R}}$ and $\boldsymbol{\tilde{y}}$. Specifically, consider the following model,
\begin{equation}
	\begin{split}
		&(\boldsymbol{\tilde{R}},\boldsymbol{\tilde{y}})\mid m, \psi_1, \psi_2\, \sim\, \prod_{q=1}^m Bern(\tilde{R}_q\mid\varphi(\psi_1))\prod_{l=1}^{m-\sum_q\tilde{R}_q}Bern(\tilde{y}_l\mid\varphi(\psi_2)),\\
		&(\psi_1,\psi_2)\mid \theta_1(\mathbf{x}),\theta_2(\mathbf{x}),\boldsymbol{\sigma^2}\,\sim\, N(\psi_1\mid \theta_1(\mathbf{x}),\sigma^2_1)N(\psi_2\mid \theta_2(\mathbf{x}),\sigma^2_2),\\
		& (\theta_1(\mathbf{x}),\theta_2(\mathbf{x}))\mid G_{\mathbf{x}}\,\sim\,  G_{\mathbf{x}},\,\,  
		G_{\mathbf{x}}\,=\,\sum_{\ell=1}^{\infty}\omega_{\ell}(\mathbf{x})\delta_{(\theta_{1\ell}(\mathbf{x}),\theta_{2\ell}(\mathbf{x}))}, 
	\end{split}
	\label{eq:hierformoverdunderly}
\end{equation}
with the same prior on $G_{\mathbf{x}}$ and $\boldsymbol{\sigma^2}$ as for the ``Gen-LNB'' model. Then, the two model formulations are equivalent in terms of an equal MGF for the respective $(R,y)$ and $(\sum\tilde{R}_q,\sum\tilde{y}_l)$. 

\begin{proposition}
\label{prop:overdequalmgf}
With the same $m$ and $G_{\mathbf{x}}$, equation (\ref{eq:equalMGF}) holds for $\mathcal{M}$ and $\tilde{\mathcal{M}}$, that is, the mixture models defined in (\ref{eq:overdgenmodelform}) and (\ref{eq:hierformoverdunderly}), respectively.
\end{proposition}

Proposition \ref{prop:overdequalmgf} allows us to implicitly condition on $m=1$ when conducting inference for the dose-response curves. We denote the logit-normal integral $\int\varphi(\psi)N(\psi\mid \theta,\sigma^2)d\psi$ by $\varepsilon(\theta,\sigma^2)$. The expressions for dose-response curves at the aforementioned three endpoints are given by
{\allowdisplaybreaks
\begin{align*}
		D(x)&\, = \, \text{Pr}(\tilde{R}=1\mid G_{\mathbf{x}},\boldsymbol{\sigma^2})\,=\,\sum_{\ell=1}^{\infty} 
\omega_{\ell}(\mathbf{x}) \, \varepsilon(\theta_{1\ell}(\mathbf{x}),\sigma_1^2);\\
    M(x)&\, = \, \text{Pr}(\tilde{y}=1\mid\tilde{R}=0,G_{\mathbf{x}},\boldsymbol{\sigma^2})
    \,=\,
     \sum_{\ell=1}^{\infty} \frac{\omega_{\ell}(\mathbf{x})
[1-\varepsilon(\theta_{1\ell}(\mathbf{x}),\sigma_1^2)]}
{\sum_{\ell=1}^{\infty}\omega_{\ell}(\mathbf{x})
[1-\varepsilon(\theta_{1\ell}(\mathbf{x}),\sigma_1^2)]} \, 
\varepsilon(\theta_{2\ell}(\mathbf{x}),\sigma_2^2);\\
    r(x)&\, = \, \text{Pr}(\tilde{R}=1 \,\, \text{or} \,\, \tilde{y}=1\mid G_{\mathbf{x}},\boldsymbol{\sigma^2})\\
    &\,=\, 1-\sum_{\ell=1}^{\infty} 
\omega_{\ell}(\mathbf{x}) \, [1-\varepsilon(\theta_{1\ell}(\mathbf{x}),\sigma_1^2)]
[1-\varepsilon(\theta_{2\ell}(\mathbf{x}),\sigma_2^2)].
\end{align*}
}
Flexible inference for the dose-response curves is again enabled with local-adjustable mixing weights.

The intracluster correlation under the general model with overdispersed kernel has a similar form as in (\ref{eq:intracorrorimodel}), in which every component should include further conditioning on $\boldsymbol{\sigma^2}$. Specifically, $\text{E}(\tilde{Y}_{qj} \mid G_{\mathbf{x}},\boldsymbol{\sigma}^2)=$ $\sum_{\ell=1}^{\infty}\omega_{\ell}(\mathbf{x})\,\{\varepsilon(\theta_{j\ell}(\mathbf{x}),\sigma_j^2)
\prod_{k=1}^{j-1}[1-\varepsilon(\theta_{k\ell}(\mathbf{x}),\sigma_k^2)]\}$, 
$\text{Var}(\tilde{Y}_{qj} \mid G_{\mathbf{x}})=$
$\text{E}(\tilde{Y}_{qj}\mid G_{\mathbf{x}})-[\text{E}(\tilde{Y}_{qj}\mid G_{\mathbf{x}})]^2$, $\forall\,q\in\{1,\ldots,m\}$. Additionally, 
 $\text{E}(\tilde{Y}_{qj}\tilde{Y}_{q^{\prime}j}\mid G_{\mathbf{x}},\boldsymbol{\sigma^2})=$
$\sum_{\ell=1}^{\infty}\omega_{\ell}(\mathbf{x})$  $\{\int \varphi^2(\psi_j)N(\psi_j\mid\theta_{j\ell}(\mathbf{x}),\sigma_j^2)d\psi_j\}\,\{\prod_{k=1}^{j-1}$ $\int[1-\varphi(\psi_k)]^2N(\psi_k\mid\theta_{k\ell}(\mathbf{x}),\sigma_k^2)d\psi_k\}$. We set $\varphi(\theta_{3\ell}(\mathbf{x}))\equiv 1$ and $\sigma_3^2=0$. The positive intracluster correlation property can also be established in this context, as the model continues to assume a shared variance/correlation within the dam.

To obtain meaningful inference, it is important to use a proper, well-calibrated prior for $\boldsymbol{\sigma^2}$. This is a challenging task because of the lack of analytical form for the logit-normal integral. Nonetheless, we propose a general strategy for specifying the $IG(a_{\sigma},b_{\sigma})$ prior, based on the approximation in Proposition \ref{prop:lnbmargapprox}, and working with the mixture kernel, i.e., the LNB distribution. From the second line of (\ref{eq:approxlnb}), and noticing $\varphi(\theta)\in(0,1)$, we can show that for small to moderate (but still providing enough variability) $\sigma^2$, the intracluster correlation is approximately $\sigma^2/4$. Simple calculation yields that modeling by LNB in lieu of Binomial provides an extra $(m-1)\sigma^2/4$ folds of the variance.
In practice, we can use a prior guess about the average variance deviation of $R$ and $y$ from the Binomial across the dose levels, and set the prior for $\boldsymbol{\sigma^2}$ accordingly, such that the overdispersion provided by the LNB kernel is enough to capture the extra variation. The other prior hyperparameters can be specified in the same fashion as the general model. Specifically, the prior specification strategy that ensures a monotonic trend in the prior expectation of dose-response curves can still be applied here.  

Another appealing feature of the proposed model comes from the posterior simulation perspective. The mixing structure of the model is inherited from the ``Gen-Bin'' model,
rendering the computational techniques developed for it readily adaptable here. We develop a blocked Gibbs sampler based on the MCMC algorithm proposed in \citet{KangKottas2022}, with modifications to account for the extra continuous mixing at the kernel. The detailed algorithm is presented in the Supporting Information. 
With the MCMC samples of model parameters, we can conduct any type of relevant inference, following the same procedure as the general model with original kernel.

To complete the spectrum of the proposed models, we also consider the simplification by removing the dose dependence in the mixing weights. That is, instead of determining weights through a LSBP prior, we use the stick-breaking formulation of the DP. We term this the common-weights mixture with product of LNB kernel (``CW-LNB'') model.

%% file: chapter4.tex
\section{Synthetic Data Examples}
\label{sec:simstudy}

We conduct simulation experiments to demonstrate the practical benefits of using nonparametric mixture models in development toxicity study. Specifically, the first experiment is designed to highlight the benefits of local, dose-dependent weights in capturing non-standard dose-response relationships. The objective of the second experiment is to illustrate the utility of the overdispersed kernel in capturing the vast heterogeneity of the data. 

\subsection{First Synthetic Data Example}
\label{subsec:sim1}

For the first experiment, we consider four active dose level at 0.625, 1.25, 2.5, and 5 $g/kg$ and a control group. We consider a total of $n=100$ dams, evenly distributed across the dose levels. For each dam, the number of implants are generated from a Poisson distribution with mean 20. Conditioning on the number of implants, the responses are generated from a three component mixture of ``CR-LNB'' model, with dose-dependent model parameters. That is, 
\begin{equation*}
    (R_{di},y_{di})\mid m_{di}\stackrel{ind.}{\sim} \sum_{k=1}^3 w_k(\mathbf{x}_d)LNB(R_{di}\mid m_{di},\theta_{1k}(\mathbf{x}_d),\sigma_1^2(\mathbf{x}_d))LNB(y_{di}\mid m_{di}-R_{di},\theta_{2k}(\mathbf{x}_d),\sigma_2^2(\mathbf{x}_d)),
\end{equation*}
where $\mathbf{x}_d=(1,x_d)$, $\theta_{jk}(\mathbf{x}_d)=b_{jk0}+b_{jk1}x_d$, for $j=1,2$ and $k=1,2,3$. The dose-dependent weights are induced by computing $p_{j}(\mathbf{x}_d)=\Phi(a_{j0}+a_{j1}x_d)$, for $j=1,2$, where $\Phi(\cdot)$ denotes the c.d.f. of the standard normal distribution, and setting $(w_1(\mathbf{x}_d),w_2(\mathbf{x}_d),w_3(\mathbf{x}_d))$ $=(p_1(\mathbf{x}_d),(1-p_1(\mathbf{x}_d))p_2(\mathbf{x}_d),(1-p_1(\mathbf{x}_d))(1-p_2(\mathbf{x}_d)))$. Additionally, $\sigma_j^2(\mathbf{x}_d)=c_{j0}+c_{j1}x_d$, for $j=1,2$, and the parameters are chosen to ensure $\sigma^2(\mathbf{x}_d)>0$. 

We visualize the simulated data set in Figure \ref{subfig:sim2data}. For each dam, we plot the observed $R_{di}/m_{di}$, $y_{di}/(m_{di}-R_{di})$, and $(R_{di}+y_{di})/m_{di}$. In each panel, the solid line indicates the true dose-response curve. We intentionally set the true dose-response curves for the malformation and the combined negative outcome endpoints to exhibit a J-shape, which is referred to as the hormetic dose-response relationship in the toxicological sciences. Hormesis is a dose–response phenomenon characterized by beneficial effect to low exposures to toxins, and thus by opposite effects in small and large doses. The posterior mean and 95\% interval estimates of dose-response curves under parametric models, i.e., the continuation-ratio logits model and the ``CR-LNB'' model, are also shown in Figure \ref{subfig:sim2data}. As expected, the standard models cannot capture the dip in the dose-response curves.  


Figure \ref{fig:sim2npcurves} displays the posterior point and interval estimates of the dose-response curves under the nonparametric mixture models. All the models considered here capture the true dose-response curve for the embryolethality endpoint well, providing point estimates that are almost identical to the true monotonically increasing function. As for the dose-response curves corresponding to the malformation and combined risk endpoints, the two nonparametric mixture models without dose-dependent mixing weights (``CW-Bin'' model and ``CW-LNB'' model) provide improved point and interval estimates comparing to their parametric backbones, but still fail in depicting the non-monotonic shape. On the contrary, the ``Gen-Bin'' model and the ``Gen-LNB'' model, permitting more efficient local adjustments, capture the dip in these dose-response curves. The comparison demonstrates the benefit of using the mixture model with dose-dependent weights, especially when the dose-response curves are expected to have non-standard shapes.

\begin{figure}[t!]
\centering
\begin{subfigure}{\textwidth}
  \centering 
  \includegraphics[width=14.8cm,height=3.7cm]{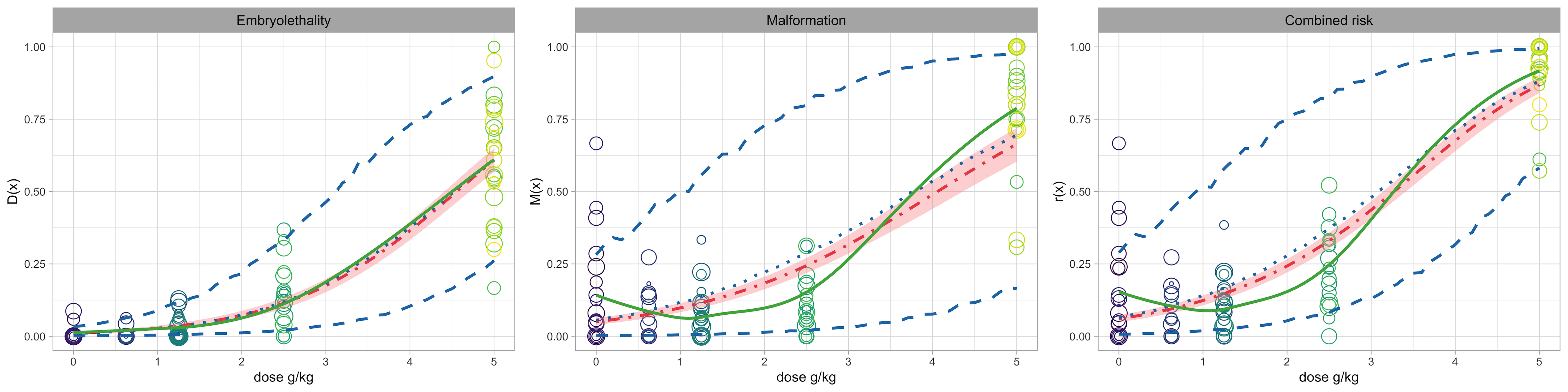}  
  \caption{\small Continuation-ratio logits model vs ``CR-LNB'' model.}
  \label{subfig:sim2data}
\end{subfigure}
\begin{subfigure}{\textwidth}
  \centering 
  \includegraphics[width=14.8cm,height=3.7cm]{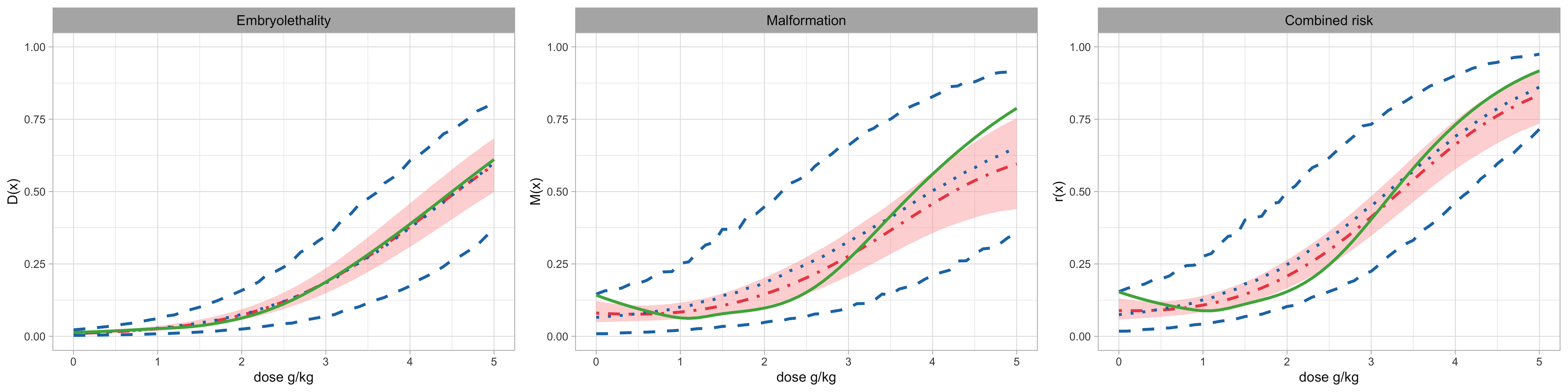}  
  \caption{\small ``CW-Bin'' model vs ``CW-LNB'' model.}
  \label{subfig:sim2cw}
\end{subfigure}
\begin{subfigure}{\textwidth}
  \centering
  \includegraphics[width=14.8cm,height=3.7cm]{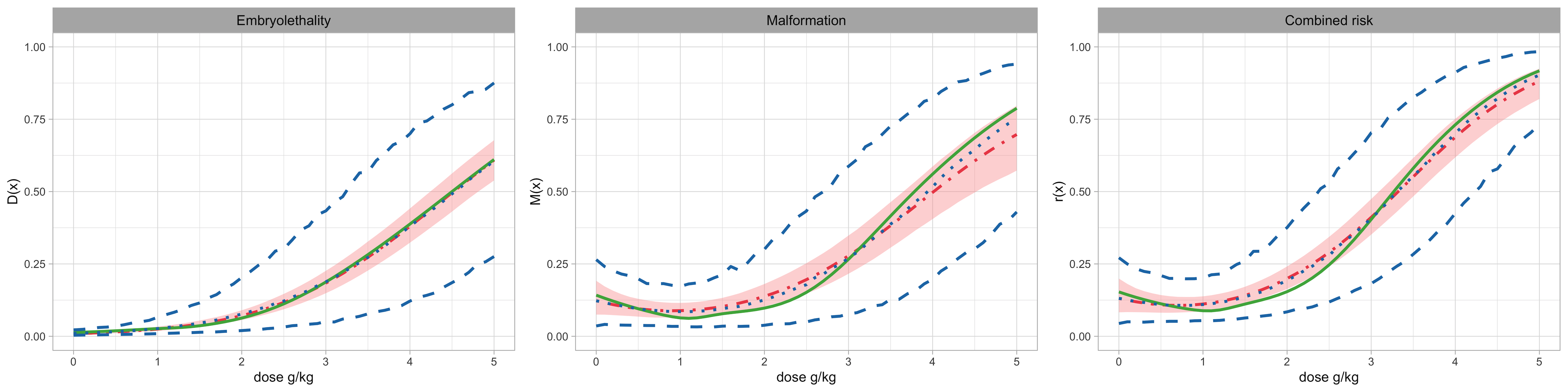}  
  \caption{\small ``Gen-Bin'' model vs ``Gen-LNB'' model.}
  \label{subfig:sim2gen}
\end{subfigure}
\caption{\small First simulation example. Posterior mean and 95\% interval estimates for the dose response curves under the mixture models with different kernel. In each panel, the posterior mean and interval estimates obtained under the model with and without overdispersed kernel are given by the blue dotted and dashed lines and 
the red dot-dashed line and shaded region, respectively. The green solid line is the true dose-response curve. In the top panel, a circle corresponds to a particular dam and the size of the circle is proportional to the number of implants.}
\label{fig:sim2npcurves}
\end{figure}

To further explore how the different nonparametric models utilize the mixture structure, Figure \ref{fig:sim2weights} shows the posterior distributions of the four largest mixture weights across dose levels. The models without overdispersed kernel generally activate more mixing components. This is to be expected, because these models rely on the mixing structure to account for overdispersion. The ``Gen-Bin'' model tends to use more mixing components at low dose region to help capture the dip of the dose-response curves. Equipped with overdispersed kernel, the ``CW-LNB'' model and the ``Gen-LNB'' model are more efficient in terms of the number of effective mixture components. Specifically, under the ``Gen-LNB'' model, we note the pronounced local adjustment of the mixing weights in the dose region where the dose-response curves change more drastically. On the contrary, the ``CW-LNB'' model does not allow for local adjustable mixture weights, and thus can not capture as effectively the non-standard local behavior of the malformation and combined risk dose-response curves. 

\begin{figure}[t!]
\centering
\includegraphics[width=14.7cm,height=4.9cm]{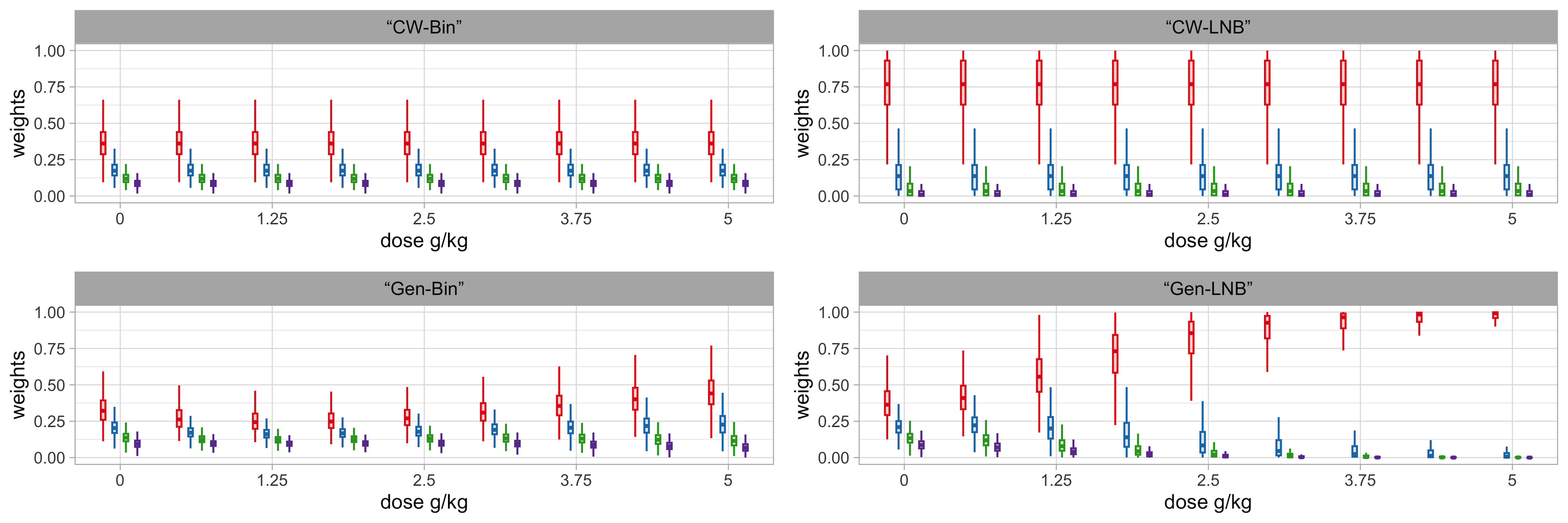}  
\caption{\small First simulation example. Box plots of the posterior samples for
the four largest mixture weights at the observed dose levels and a new dose level, under each of the nonparametric models.}
\label{fig:sim2weights}
\end{figure}

\subsection{Second Synthetic Data Example}
\label{subsec:sim2}

We consider active dose levels at 0.625, 1.25, 2.5, 3.75, and 5 $g/kg$, and a control group. A total of $n=150$ dams are randomly assigned across the dose levels with uniform probability. We adopt the same process with the first simulation example to generate the number of implants for each dam. Then, the ordinal responses are obtained by sampling from
\begin{equation*}
    (R_{di},y_{di})\mid m_{di}\stackrel{ind.}{\sim} BB(R_{di}\mid m_{di},\theta_{1}(\mathbf{x}_d),\lambda_1(\mathbf{x}_d))BB(y_{di}\mid m_{di}-R_{di},\theta_{2}(\mathbf{x}_d),\lambda_2(\mathbf{x}_d)),
\end{equation*}
where $\theta_j(\mathbf{x}_d)=b_{j0}+b_{j1}x_d$, and $\lambda_j(\mathbf{x}_d)=c_{j0}+c_{j1}x_d>0$, for $j=1,2$.  
The data are visualized in Figure \ref{subfig:sim3data}, including the posterior point and 95\% interval estimates under the continuation-ratio logits model and the ``CR-BB'' model. Although the true dose-response curves here have relatively standard increasing shape, the parametric models suffer in uncertainty quantification, due to the vast heterogeneity of the data. In particular, the ``CR-BB'' model is similar to the true data generating process, but the interval estimates obtained under it are too wide to be practically useful. 


The nonparametric mixture models are applied to the data. Figure \ref{fig:sim3npcurves} plots posterior point and interval estimates for the dose-response curves. The nonparametric mixture models behave comparably in terms of recovering the underlying regression curves, evidenced by the similar posterior mean for the dose-response curves. The interval estimates under all the models capture the true dose-response curves. As expected, models with overdispersed kernel result in wider posterior uncertainty bands than the models with the continuation-ratio logits kernel. Observing the extensive dispersion in the data, a wider uncertainty band is arguably more reasonable. 

\begin{figure}[t!]
\centering
\begin{subfigure}{\textwidth}
  \centering 
  \includegraphics[width=14.8cm,height=3.7cm]{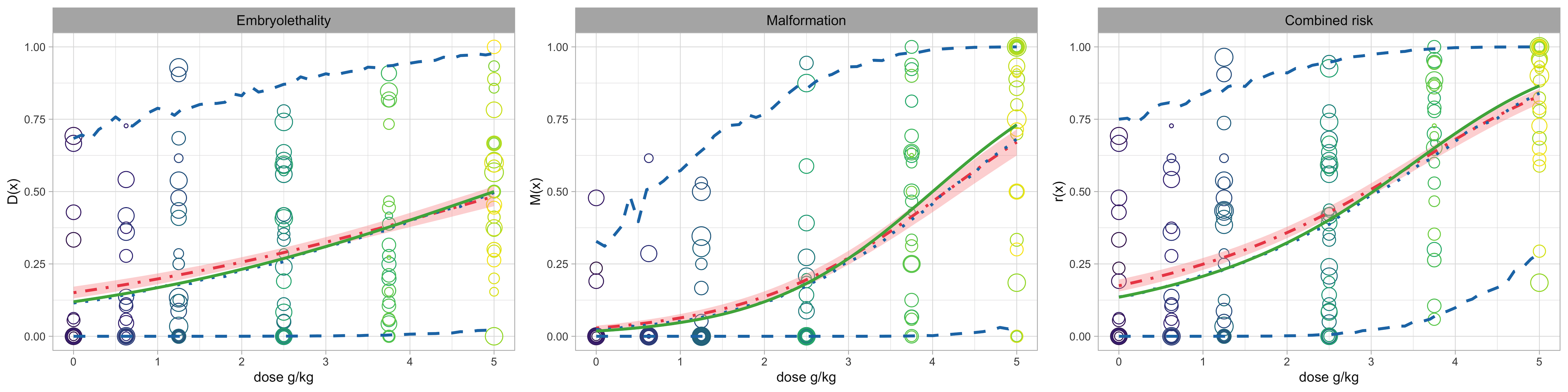}  
  \caption{\small Continuation-ratio logits model vs ``CR-BB'' model.}
  \label{subfig:sim3data}
\end{subfigure}
\begin{subfigure}{\textwidth}
  \centering 
  \includegraphics[width=14.8cm,height=3.7cm]{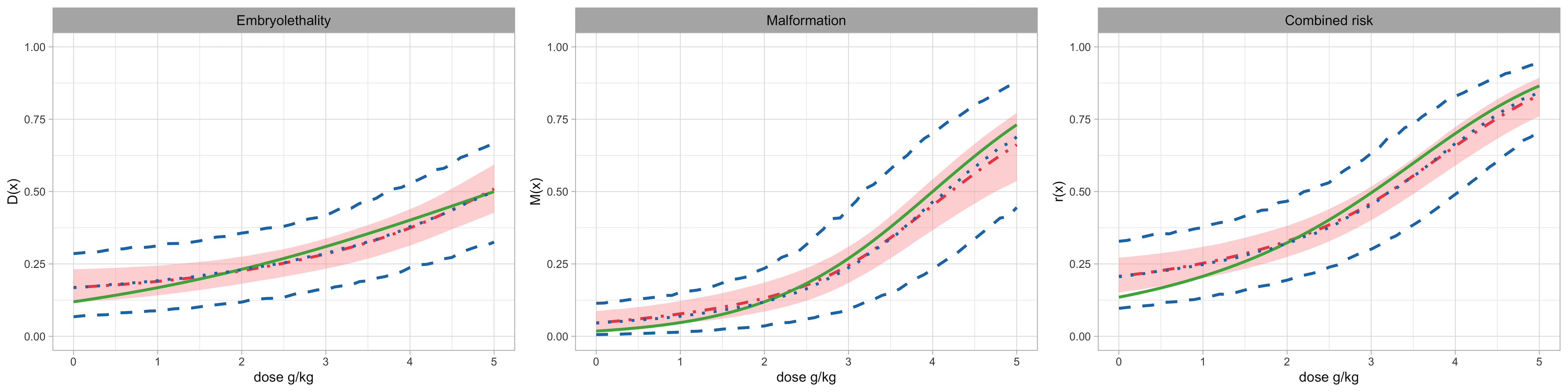}  
  \caption{\small ``CW-Bin'' model vs ``CW-LNB'' model.}
  \label{subfig:sim3cw}
\end{subfigure}
\begin{subfigure}{\textwidth}
  \centering
  \includegraphics[width=14.8cm,height=3.7cm]{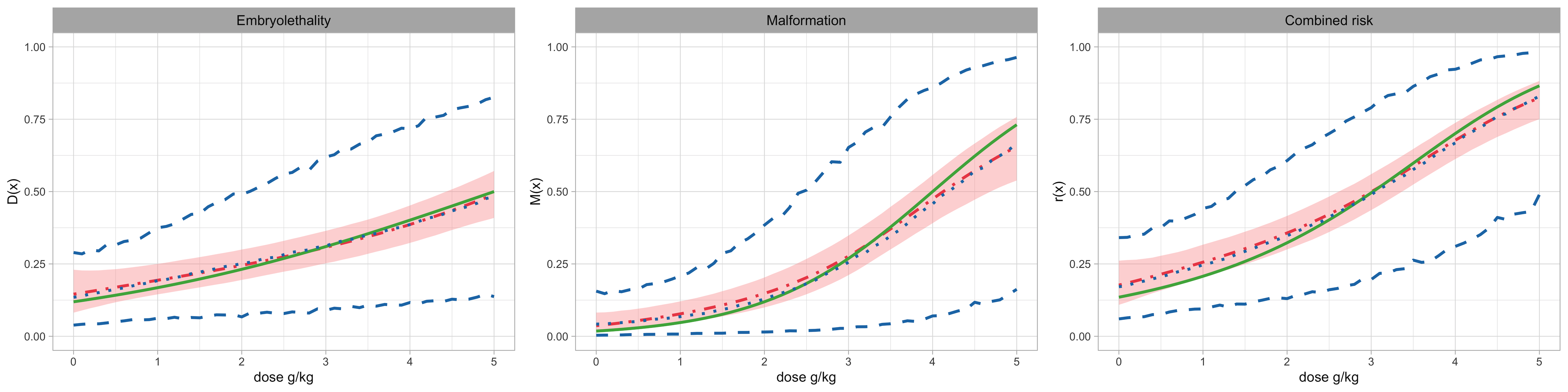}  
  \caption{\small ``Gen-Bin'' model vs ``Gen-LNB'' model.}
  \label{subfig:sim3gen}
\end{subfigure}
\caption{\small Second simulation example. Posterior mean and 95\% interval estimates for the dose response curves under the mixture models with different kernel. In each panel, the posterior mean and interval estimates obtained under the model with and without overdispersed kernel are given by the blue dotted and dashed lines and 
the red dot-dashed line and shaded region, respectively. The green solid line is the true dose-response curve. In the top panel, a circle corresponds to a particular dam and the size of the circle is proportional to the number of implants.}
\label{fig:sim3npcurves}
\end{figure}

Moreover, we assess the performance of these nonparametric mixture model in capturing the overdispersion. We obtain the posterior distribution of the intracluster correlation $\text{Corr}(\tilde{Y}_{qj},\tilde{Y}_{q^{\prime}j}\mid G_{\mathbf{x}})$, for $j=1,2,3$, under each model and compare it with the truth. We also conduct sensitivity analysis regarding the prior of the overdispersion parameter. These analyses are shown in the Supporting Information. 

%% file: chapter5.tex
\section{Real Data Illustration}
\label{sec:dtdataill}

Working with the EG data, we illustrate the four nonparametric mixture models in addressing a spectrum of risk assessment tasks. We work with the (conservative) truncation level of $L=50$ for the blocked Gibbs sampler. Posterior inference relies on 5000 MCMC samples, taken every 2 iterations from a chain of 30000 iterations, with a burn-in period of 20000.

We set the prior hyperparameters such that a monotonic increasing trend is incorporated in the prior expected dose-response curves. In line with this objective, the key is to specify $\boldsymbol{\mu}_{0j}$ and $\Lambda_{0j}$, $j=1,2$, following the strategy proposed in \citet{KangKottas2022}. The hyperparameters regarding the mixing weights under either the common-weights mixture or the general mixture are set such that they favor a priori a comparable number of distinct components. We use $IG(3,1.2)$ as the prior for the overdispersion parameters in $\boldsymbol{\sigma^2}$, which provides approximately an extra $15\%$ variance on average. For prior sensitivity analysis, we assume an alternative, more diffused prior, that is, $IG(2,0.6)$. Despite the choice of prior and mixing structure, the posterior distributions of $\sigma_1^2$ and $\sigma_2^2$ are comparable.

\begin{figure}[t!]
\centering
\begin{subfigure}{\textwidth}
  \centering 
  \includegraphics[width=14.8cm,height=3.7cm]{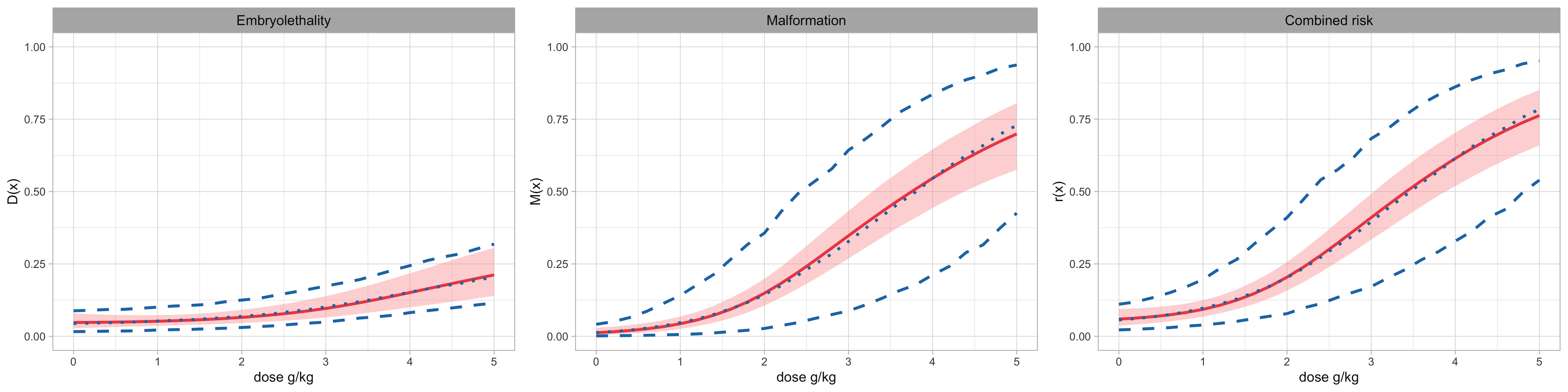}  
  \caption{\small ``CW-Bin'' model vs ``CW-LNB'' model.}
  \label{subfig:egcw}
\end{subfigure}
\begin{subfigure}{\textwidth}
  \centering
  \includegraphics[width=14.8cm,height=3.7cm]{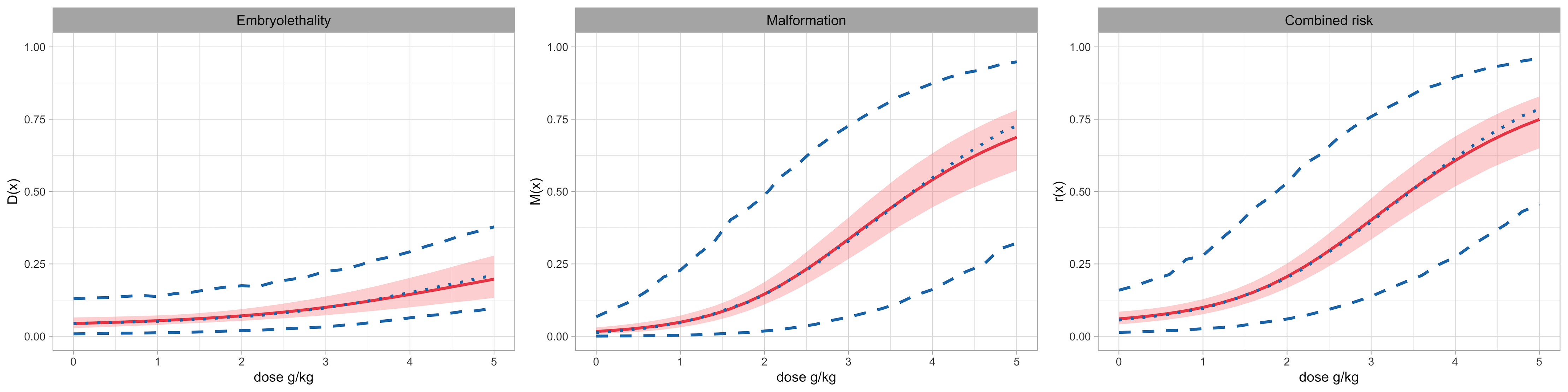}  
  \caption{\small ``Gen-Bin'' model vs ``Gen-LNB'' model.}
  \label{subfig:eggen}
\end{subfigure}
\caption{\small EG data. Posterior mean and 95\% interval estimate for the dose response curves under the mixture models with different kernel. In each panel, the posterior mean
and interval estimates obtained under the model with and without overdispersed kernel are given
by the blue dotted and dashed lines and the red dot-dashed line and shaded region, respectively. }
\label{fig:egnpcurves}
\end{figure}

Posterior estimates of dose-response curves under the discrete mixture models are displayed in Figure \ref{fig:egnpcurves}. The difference among the posterior point estimates of the dose-response curves is minor. The uncertainty bands provided by the models with overdispersed kernel are significantly wider, with the width changing across toxin levels. As shown in Figure \ref{fig:plotdata}, the variability of the responses increases with dose level. Uncertainty bands obtained under the ``CW-Bin'' and ``Gen-Bin'' model capture the trend in general, while they seem to underestimate the influence of the dose levels. Moreover, illustrated by a wider interval compared to that at 5 g/kg, the models with overdispersed kernel intensify the uncertainty at the region from 3 g/kg to 4 g/kg, where interpolation is actually needed. Without the help of overdispersed kernel, the models tend to be overconfident at this region. 
We also notice that comparing with the estimates under the continuous mixture model (Figure \ref{fig:egparam}), the uncertainty bands obtained here are more plausible, indicating more effective control of variability under the discrete nonparametric mixture models.      


\begin{figure}[t!]
\centering
\includegraphics[width=14.8cm,height=3.7cm]{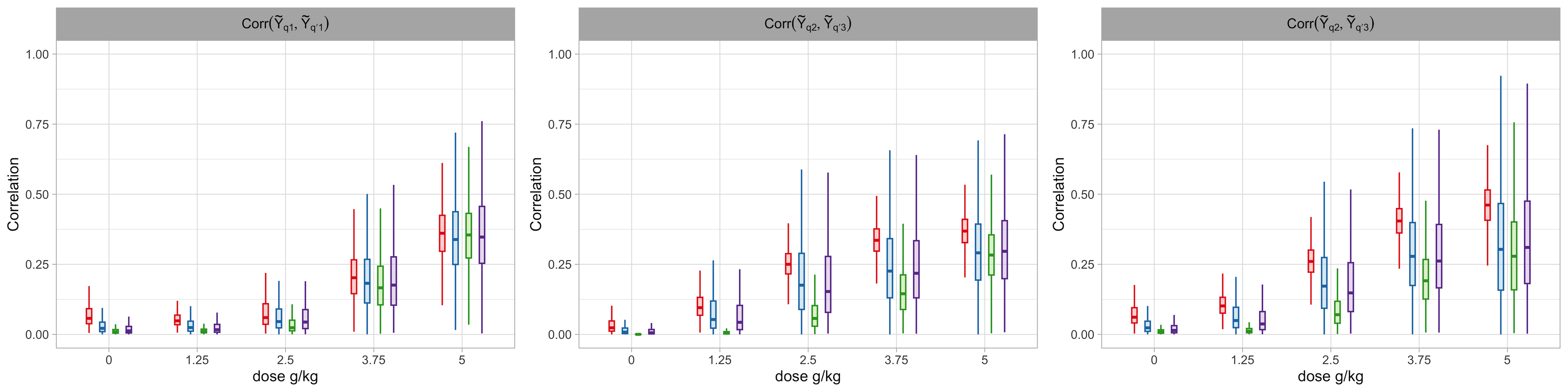}
\caption{\small EG data. Box plot of the intracluster correlation posterior distributions at four observed toxin levels and for the new value of $x=3.75$ g/kg. In each panel, estimates under ``CW-Bin'', ``CW-LNB'', ``Gen-Bin'' and ``Gen-LNB'' model are shown in red, blue, green, and purple, respectively.}
\label{fig:egcorr}
\end{figure}

We display posterior samples of the intracluster correlations at the observed toxin levels and a new level $x=3.75$ g/kg, with the box plots in Figure \ref{fig:egcorr}. Despite the model and endpoint, the correlations depict an increasing trend with toxin levels, consistent with the observed data pattern. 
Moreover, the intracluster correlation at the new dose level is approximately the average of the correlations at the two observed neighbors, indicating a smooth borrowing of strength across dose levels.
As expected, the distribution of correlations from models with overdispersed kernel spread a wider range.  
Also noteworthy is that 
the magnitude of the intracluster correlation under the ``CW-Bin'' model is generally larger than the other models, which also means a larger variance for the response. However, as shown in Figure \ref{subfig:egcw}, the posterior uncertainty for the dose-response curve under the ``CW-Bin'' model is shorter. This incongruity indicates a weak control of variability under the ``CW-Bin'' model, in which both the mixing structure and the kernel are restricted.

Estimating the effective dose (ED) and the benchmark dose (BMD) is crucial in developmental toxicity risk assessment. The procedure initiates with specifying the benchmark response level (BMR), denoted by $\alpha$. After a dose-response model is applied to the data, the effective dose $\text{ED}_{\alpha}$ is defined as the dose that induces an excess risk of $\alpha$ over control. As an example, for the embryolethality endpoint, the ED is found as the solution to the equation $D(\text{ED}_{\alpha}^{D})-D(0)/(1-D(0))=\alpha$. With posterior samples of $D(x)$, we can numerically solve the equation, and obtain the posterior distribution of $\text{ED}_{\alpha}^{D}$. Analogously, we obtain the posterior distribution of ED corresponds to the malformation and combined risk endpoints, using posterior realizations of $M(x)$ and $r(x)$, respectively. Then, $\text{BMD}_{\alpha}$ is defined as the left endpoint of the 95\% credible interval of $\text{ED}_{\alpha}$.  \citet{Allen1994} found that BMD with $\alpha=5\%$ is similar to the no observed adverse effect level (NOAEL). Additionally, agencies recommend reporting BMD at the level of $10\%$ extra risk for dichotomous data \citep{EPA2012}. We focus on obtaining the posterior distribution of ED, and estimating BMD, at the three endpoints, for $\alpha=5\%$ and $\alpha=10\%$.

\begin{figure}[t!]
\centering
\begin{subfigure}{\textwidth}
  \centering 
  \includegraphics[width=14.8cm,height=3.7cm]{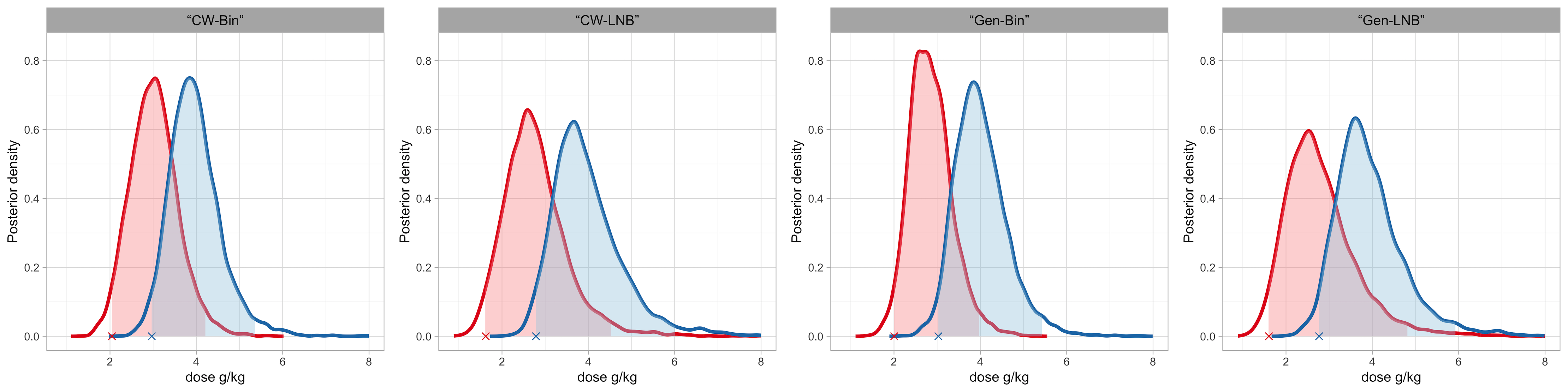}  
  \caption{\small Embryolethality endpoint.}
  \label{subfig:eged1df}
\end{subfigure}
\begin{subfigure}{\textwidth}
  \centering
  \includegraphics[width=14.8cm,height=3.7cm]{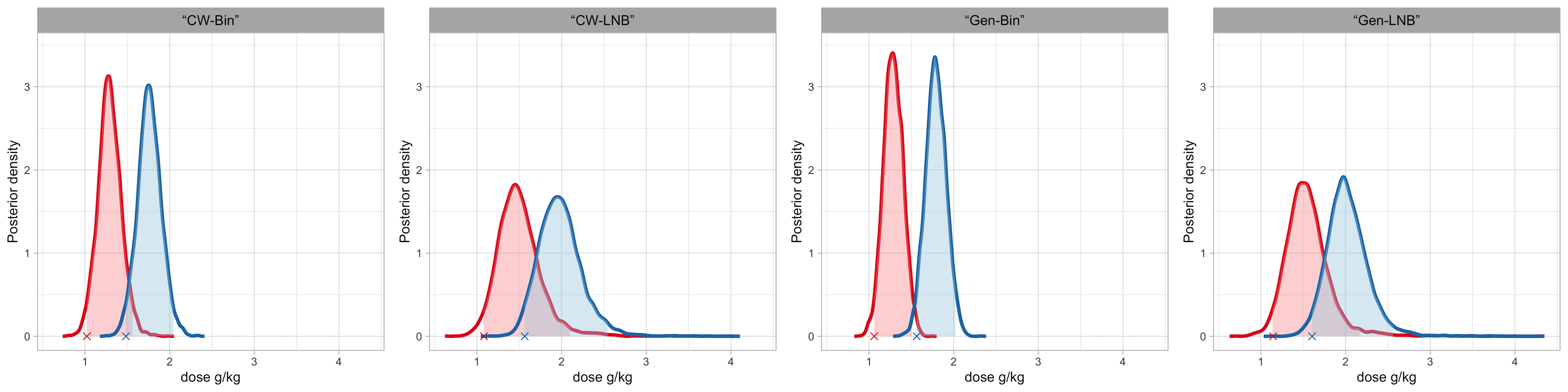}  
  \caption{\small Malformation endpoint.}
  \label{subfig:eged2df}
\end{subfigure}
\begin{subfigure}{\textwidth}
  \centering
  \includegraphics[width=14.8cm,height=3.7cm]{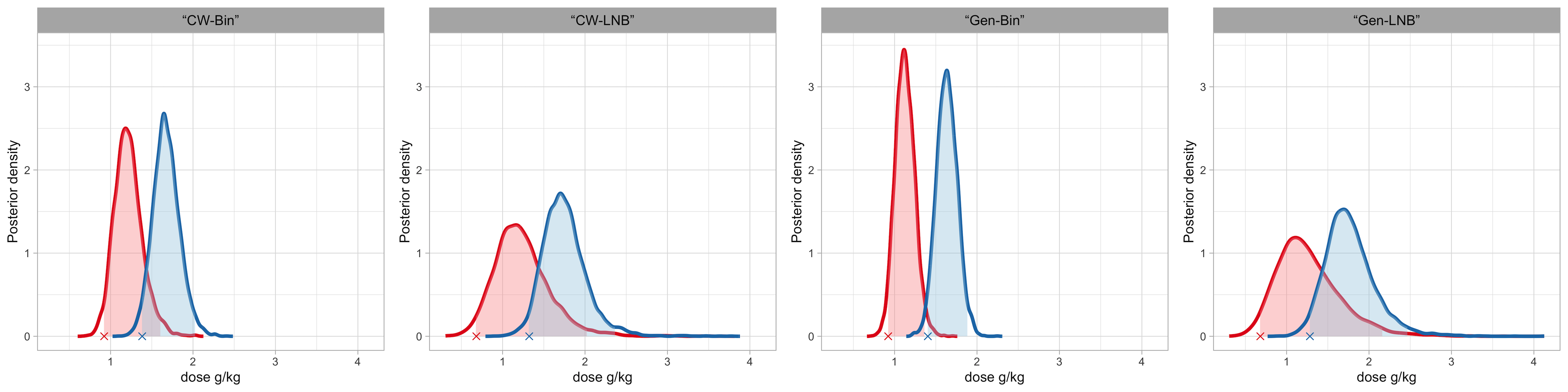}  
  \caption{\small Combined risk endpoint.}
  \label{subfig:eged3df}
\end{subfigure}
\caption{\small EG data. Posterior distribution of the effective dose with 5\% BMR (in red) and 10\% BMR (in blue). The shaded region indicates the 95\% credible interval. The corresponding benchmark dose is marked with ``$\times$''. }
\label{fig:egdosefind}
\end{figure}

Figure \ref{fig:egdosefind} plots the posterior distribution of ED and the estimated BMD. We note that for embryolethality endpoint, the posterior samples of ED include extrapolation of toxin levels. The models with overdispersed kernel yield more dispersed distributions, while the models with the same kernel provide comparable results, despite the mixing structure. The estimated BMDs are summarized in Table \ref{tab:egbmd}. Results are generally robust across models. 
Therefore, it is manifested that the models themselves, in the absence of incorporating biochemical characteristics, are adequate for estimating BMD.  

\begin{table}[t!] \centering
\small
\caption{\small EG data. BMD estimation under different models, based on posterior samples of ED.} 
\label{tab:egbmd}
\begin{tabular}{ccccccc}
\hline
\hline
\multirow{2}{*}{Model} & \multicolumn{2}{c}{Embryolethality} & \multicolumn{2}{c}{Malformation} & \multicolumn{2}{c}{Combined risk} \\
\cline{2-7} & $\alpha=5\%$ & $\alpha=10\%$ & $\alpha=5\%$ & $\alpha=10\%$ & $\alpha=5\%$ & $\alpha=10\%$\\ 
\hline
``CW-Bin'' &   2.05 & 2.97 & 1.02 & 1.48 & 0.92 & 1.38\\
``CW-LNB'' &   1.62 & 2.79 & 1.08 & 1.56 & 0.68 & 1.32\\
``Gen-Bin'' &  2.00 & 3.03 & 1.06 & 1.56 & 0.92 & 1.40\\
``Gen-LNB'' &  1.64 & 2.77 & 1.14 & 1.60 & 0.68 & 1.28\\
\hline
\hline
\end{tabular}
\end{table}

\begin{figure}[t!]
\centering
\includegraphics[width=14.8cm,height=7.4cm]{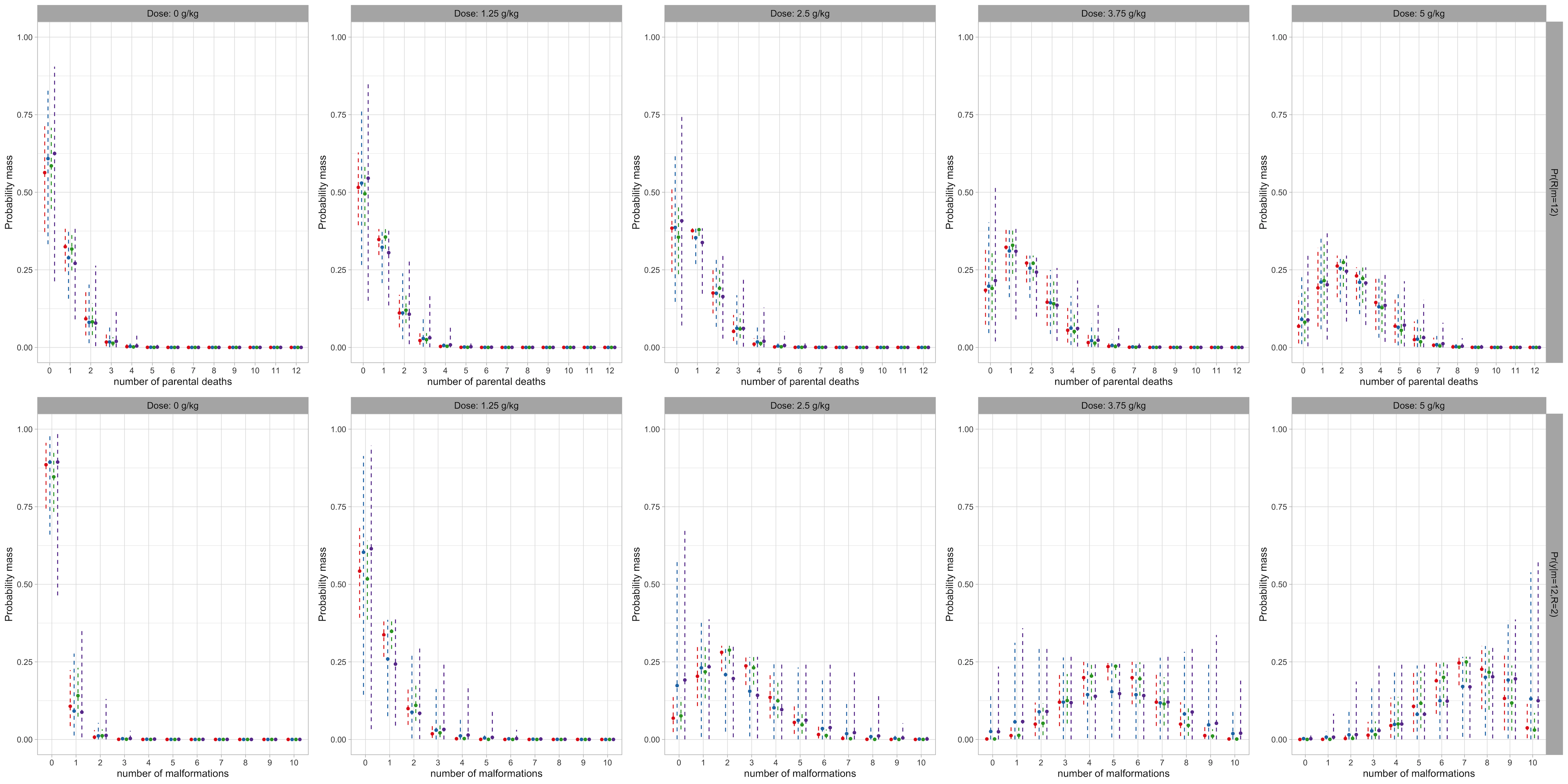}
\caption{\small EG data. Posterior mean (``$\circ$'') and $95\%$ uncertainty bands
(dashed lines) for the probability mass $\text{Pr}(R\mid m=12,G_{\mathbf{x}})$ (top panels) and conditional probability mass $\text{Pr}(y\mid m=12,R=2,G_{\mathbf{x}})$ (bottom panels), at four observed toxin levels and for the new value of $x=3.75$ g/kg. In each panel, estimates under ``CW-Bin'', ``CW-LNB'', ``Gen-Bin'' and ``Gen-LNB'' model are shown in red, blue, green, and purple, respectively.}
\label{fig:egpostpredpm}
\end{figure}

Figure \ref{fig:egpostpredpm} displays estimates for the probability mass functions corresponding to the number of non-viable fetuses given a specific number of implants 
$\text{Pr}(R\mid m=12,G_{\mathbf{x}})$, and the conditional probability mass functions of the number of malformations given a specified number of implants and the associated number of non-viable fetuses $\text{Pr}(y\mid m=12,R=2,G_{\mathbf{x}})$. All the models can uncover non-standard distributional shapes, especially for high toxin levels.  
Also noteworthy is the smooth evolution from right to left skewness in the conditional probability mass functions as the toxin level increases.

\begin{figure}[t!]
\centering
\begin{subfigure}{0.485\textwidth}
  \centering 
  \includegraphics[width=7.3cm,height=2.43cm]{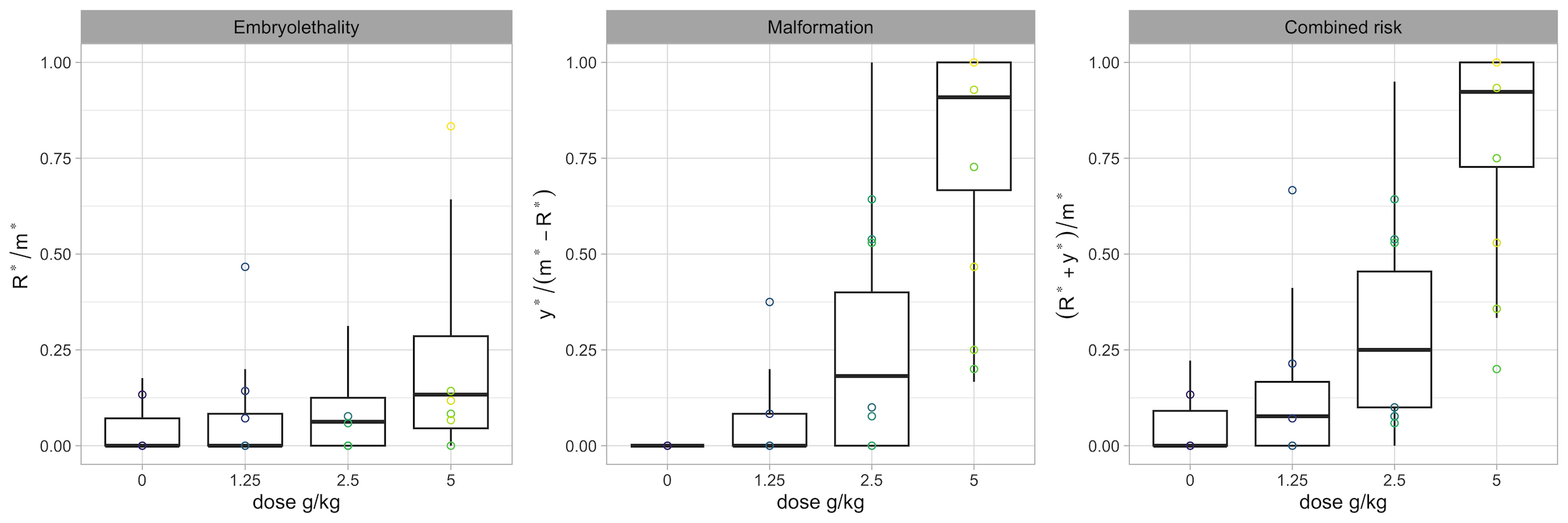}  
  \caption{\small ``CW-Bin'' model.}
  \label{subfig:egcwcheck}
\end{subfigure}
\hfill
\begin{subfigure}{0.485\textwidth}
  \centering
  \includegraphics[width=7.3cm,height=2.43cm]{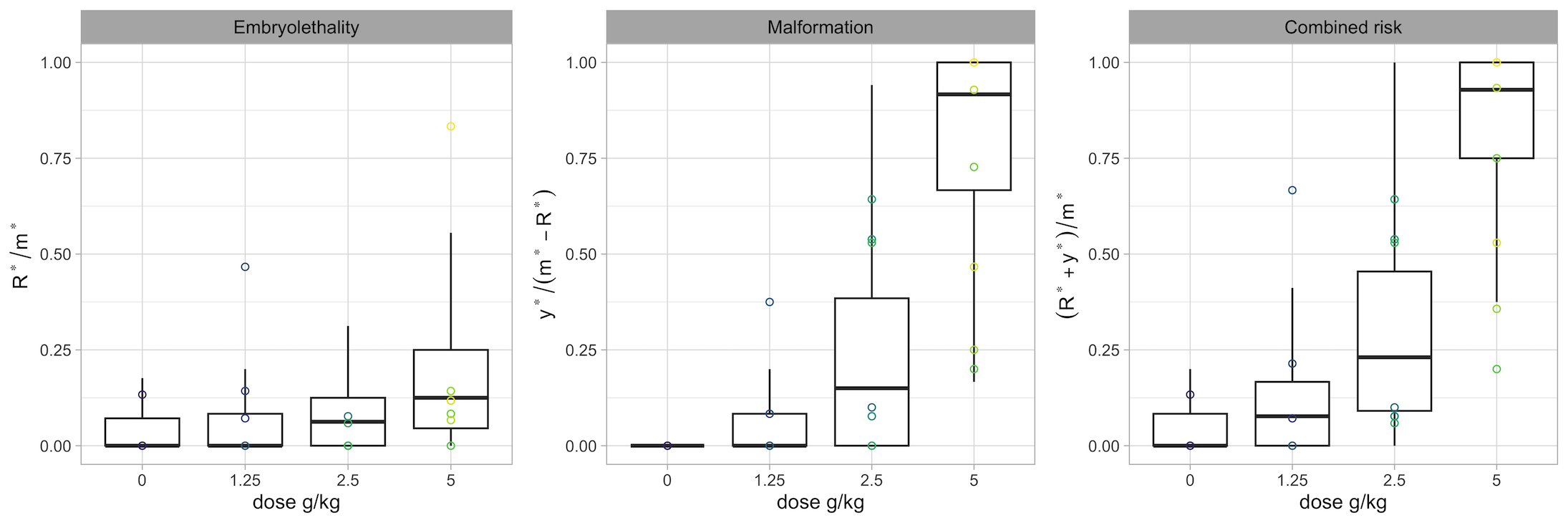}  
  \caption{\small ``CW-LNB'' model.}
  \label{subfig:egcwodcheck}
\end{subfigure}
\vskip\baselineskip
\begin{subfigure}{0.485\textwidth}
  \centering 
  \includegraphics[width=7.3cm,height=2.43cm]{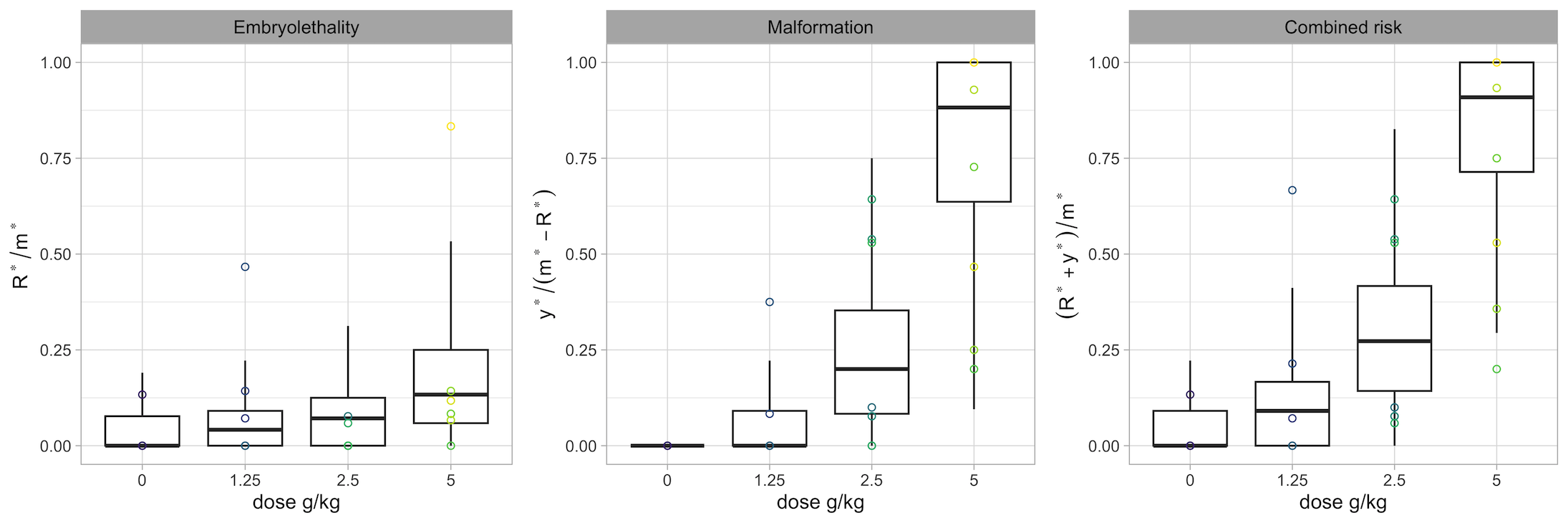}  
  \caption{\small ``Gen-Bin'' model.}
  \label{subfig:eggencheck}
\end{subfigure}
\hfill
\begin{subfigure}{0.485\textwidth}
  \centering
  \includegraphics[width=7.3cm,height=2.43cm]{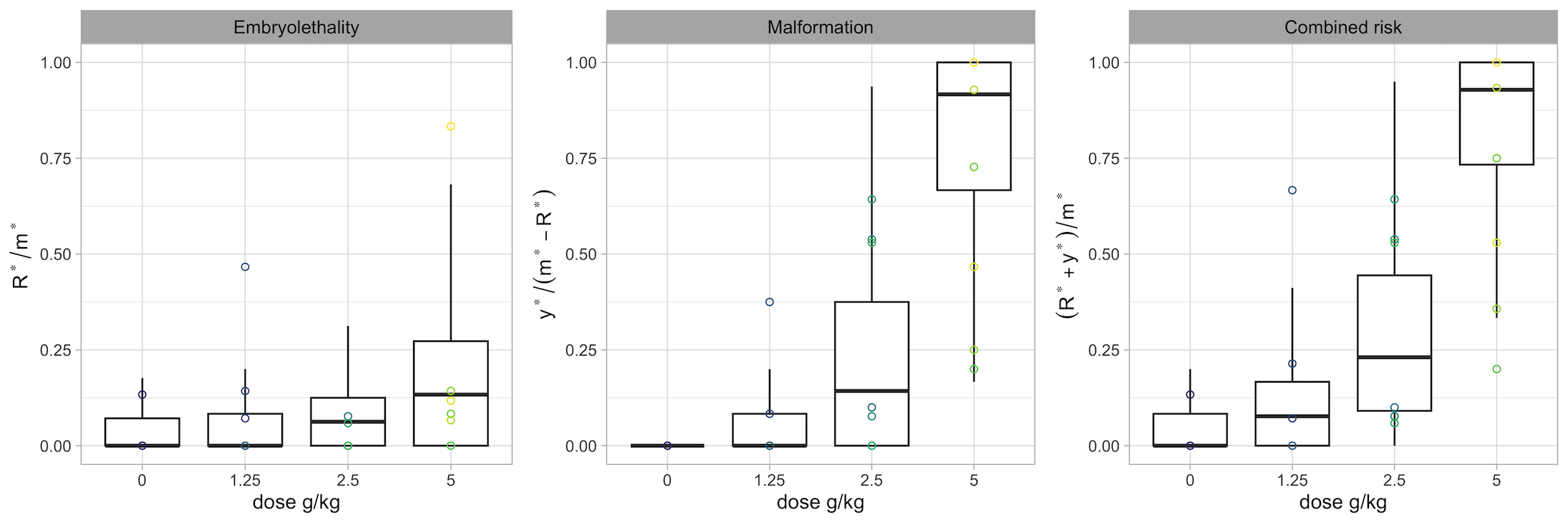}  
  \caption{\small ``Gen-LNB'' model.}
  \label{subfig:eggenodcheck}
\end{subfigure}
\caption{\small EG data. Box plots of posterior predictive samples for the embryolethality, malformation, and combined risk endpoints at the observed toxin levels. The corresponding observed proportions are denoted by ``$\circ$''. }
\label{fig:egcheck}
\end{figure}

For the models considered in this section, we preform posterior predictive model checking based on cross-validation. Specifically, we use one randomly chosen sample comprising data from 22 dams (approximately $20\%$ of the data) spread roughly evenly across the dose levels as the test set, denoted by $\{(m_{di}^{\prime},R_{di}^{\prime},y_{di}^{\prime}): d=1,\ldots,N, i=1,\ldots,n_d^{\prime}\}$. After fitting each model to the reduced data, we obtain, at each MCMC iteration, one set of posterior predictive sample at each observed dose level, denoted as $m_d^*$, $R_d^*$, and $y_d^*$. This is because the responses from the $n_d$ dams at the $d$-th dose level share the same covariate $\mathbf{x}_d$. Based on the posterior predictive samples, Figure \ref{fig:egcheck} displays box plots of the proportion of embryolethality $R^*_d/m^*_d$, malformation among live pups $y^*_d/(m^*_d-R^*_d)$, and combined negative outcomes $(R^*_d+y^*_d)/m_d^*$. The observed proportions from the test data points are marked by circles. None of these figures show evidence of ill-fitting.  

We consider two types of model comparison, based on either the posterior predictive loss (PPL) criterion \citep{GelfandGhosh1998} or the interval score (IS) criterion \citep{Gneiting2007}. The PPL criterion focuses on the first two moments of the predictive distribution, which may not be comprehensive for risk assessment in developmental toxicity studies. As a complement, the IS criterion considers the quantiles of the predictive distribution, providing a more comprehensive perspective of the predictive distribution. 
We conduct model comparison based on these criteria applied to each of the endpoints. 
For instance, consider the embryolethality endpoint, we define the goodness-of-fit term as $G(\mathcal{M})=\sum_{d=1}^N\sum_{i=1}^{n_d^{\prime}}\{R^{\prime}_{di}/m^{\prime}_{di}-\text{E}(R_d^*/m_d^*\mid\text{data})\}$, and the penalty term for model complexity as $P(\mathcal{M})=\sum_{d=1}^Nn_{d}^{\prime}\text{Var}(R_d^*/m_d^*\mid\text{data})$. 
The PPL criterion, comprised by these two terms, favors the model $\mathcal{M}$ that minimizes them. The IS criterion regarding the $95\%$ credible interval is given by 
\begin{equation*}
	S(\mathcal{M})=\sum_{d=1}^N\sum_{i=1}^{n_d^{\prime}}\{(u_d^e-l_d^e)+\frac{2}{\alpha}(l_d^e-\frac{R^{\prime}_{di}}{m^{\prime}_{di}})\mathbf{1}(\frac{R^{\prime}_{di}}{m^{\prime}_{di}}<l_d^e)+\frac{2}{\alpha}(\frac{R^{\prime}_{di}}{m^{\prime}_{di}}-u_d^e)\mathbf{1}(\frac{R^{\prime}_{di}}{m^{\prime}_{di}}>u_d^e)\},
\end{equation*}
where $l_d^e$ and $u_d^e$ denote the lower and upper limit of the posterior predictive $95\%$ credible interval of the embryolethality endpoint at dose $x_d$, respectively, and $\alpha=5\%$. The model with the smallest $S(\mathcal{M})$ is preferred. These terms are defined analogously for the other two endpoints, based on posterior predictive samples $y^*_d/(m^*_d-R^*_d)$ and $(R^*_d+y^*_d)/m_d^*$. We report the results in Table \ref{tab:egmodelcomp}.

\begin{table}[t!] \centering
\caption{\small EG data. Summary of comparison among the nonparametric models using the posterior predictive loss and interval score criteria. The values in bold correspond to the model favored by the particular criterion.} 
\label{tab:egmodelcomp}
\begin{tabular}{cccccc}
\hline
\hline
Endpoint & Criterion & ``CW-Bin'' & ``CW-LNB'' & ``Gen-Bin'' & ``Gen-LNB'' \\
\hline
\multirow{3}{*}{Embryolethality} & $G(\mathcal{M})$  & 0.72 & 0.72 & \textbf{0.71} & 0.72\\
& $P(\mathcal{M})$ & 0.56 & 0.53 & \textbf{0.45} & 0.58 \\
& $S(\mathcal{M})$ & 20.73 & \textbf{18.45} & 20.46  & 18.73 \\
\hline
\multirow{3}{*}{Malformation} & $G(\mathcal{M})$ & 1.34 & 1.39 & \textbf{1.33} & 1.36\\
& $P(\mathcal{M})$  & 1.18 & 1.10 & \textbf{0.95} & 1.17 \\
& $S(\mathcal{M})$  & 16.07    & 14.97  & 16.81   & \textbf{14.93} \\  
\hline
\multirow{3}{*}{Combined risk} & $G(\mathcal{M})$ & 1.46 & 1.50 & \textbf{1.43} & 1.49\\
& $P(\mathcal{M})$ & 1.08 & 1.01 & \textbf{0.89} & 1.03\\
& $S(\mathcal{M})$ & 25.84  & 23.50  & 27.11  & \textbf{20.91}\\
\hline
\hline
\end{tabular}
\end{table}

Based on the PPL criterion, the ``Gen-Bin'' model is preferred. The two models with overdispersed kernel have comparable goodness-of-fit, while as expected, have large penalty terms as well. Interestingly, the ``CW-Bin'' model also yields large penalty term. This is because the model imposes the most restricted structure, and thus tends to activate a larger number of effective components to capture the heterogeneity of the data. The IS criterion suggests an improvement from the use of mixture models with overdispersed kernel. In particular, for the malformation and combined risk endpoints, which exhibit vast variability, the  ``Gen-LNB'' model that allows for the highest level of flexibility is preferred. 

%% file: chapter6.tex
\section{Summary and Remarks}
\label{sec:sumrmks}

We have explored modeling approaches for developmental toxicology data, with the 
focus on introducing overdispersion through a mixing structure. We first illustrate that popular 
continuous mixture models are limited in terms of reliable uncertainty quantification for the 
dose-response curves. Contrarily, discrete mixture models, with mixing structure induced by 
dose-dependent stick-breaking process priors, offer several practical benefits. Notably, the 
enhanced flexibility results in rich inference for the response distributions and for the dose-response 
relationships. Pursuing a more effective control of variability, we consider combining the two types 
of mixture models. Specifically, the general model is formulated with a continuous mixture model 
as the kernel in a discrete nonparametric mixing structure. We show that the derived models inherit 
the properties of their backbones, while ensuring efficient posterior inference. Data from a 
toxicity experiment involving an organic solvent were used to illustrate the discrete mixture 
models and to compare their performance with regard to a series of risk assessment tasks.

A practical aspect entails selecting the appropriate model for a given problem. 
The EG data example presented illuminates a plausible avenue. 
The ``CW-Bin'' model imposes restrictions in both the mixing kernel and the mixing weights, 
and may struggle with data with vast heterogeneity. If the risk assessment task involves only 
the first two moments of the predictive distribution, the ``Gen-Bin'' model may be more suitable. 
The key advantage of incorporating overdispersed kernel within a nonparametric mixture model lies 
in improved posterior predictive interval estimation. 
Among the two models with overdispersed kernel, the ``Gen-LNB'' model offers the most flexible 
structure for overdispersion, especially helpful if the data exhibit extensive variability. 
We note that, in order to obtain effective risk assessment in developmental toxicity studies, 
a comprehensive exploration of possible modeling options is advocated by regulatory agencies.
The proposed modeling framework may be viewed as a contribution in that direction.

%
%

%% file: Supplementary.tex
\pagebreak
\vspace*{50pt}
\begin{center}
\textbf{\LARGE Supporting Information: Bayesian Nonparametric 
Risk Assessment in Developmental Toxicity Studies with Ordinal Responses}
\end{center}
\bigskip
\bigskip
\bigskip
\bigskip
\setcounter{equation}{0}
\setcounter{figure}{0}
\setcounter{table}{0}
\setcounter{page}{1}
\setcounter{section}{0}
\makeatletter
\renewcommand{\theequation}{A\arabic{equation}}
\renewcommand{\thefigure}{A\arabic{figure}}
\renewcommand{\thetable}{A\arabic{table}}
\renewcommand{\bibnumfmt}[1]{[A#1]}
\renewcommand{\citenumfont}[1]{A#1}



\section{MCMC Posterior Simulation Details}
\label{sec:smmcmcdetail}

In this section, we design the posterior simulation steps for the ``Gen-LNB'' model, and discuss its modification to accommodate for the ``CW-Bin'' model. Consider the data $\{(x_d,\mathbf{Y}_{di}): d=1,\ldots,N;i=1,\ldots,n_d\}$. For the continuous mixing structure in the kernel, and the enveloping discrete mixing structure, we introduce latent variables $\{\boldsymbol{\psi}_{di}=(\psi_{di1},\psi_{di2})\}$, and configuration variables $\{\mathcal{L}_{di}\}$. The model is formulated hierarchically as

\begin{equation*}
\begin{array}{rcl}
(R_{di},y_{di}) \mid \boldsymbol{\psi}_{di} & \stackrel{ind.}{\sim} & Bin(R_{di}\mid m_{di},\varphi(\psi_{di1}))Bin(y_{di}\mid m_{di}-R_{di},\varphi(\psi_{di2}))\\
\boldsymbol{\psi}_{di} \mid \{ \boldsymbol{\beta}_{j \ell} \},\mathcal{L}_{di}, \boldsymbol{\sigma^2} & \stackrel{ind.}{\sim} & \prod\limits_{j=1}^2 N(\psi_{dij}\mid \mathbf{x}_{d}^T\boldsymbol{\beta}_{j\mathcal{L}_{di}},\sigma^2_j), \,\,\, d=1,\ldots,N \,\,\, i=1,\ldots,n_d \\ 
\mathcal{L}_{di} \mid \{ \boldsymbol{\gamma}_{\ell} \} & \stackrel{ind.}{\sim} &  
\sum\limits_{\ell=1}^L p_{d\ell} \, \delta_{\ell}(\mathcal{L}_i), \,\,\, d=1,\ldots,N \,\,\, i=1,\ldots,n_d
\\
\boldsymbol{\beta}_{j\ell} \mid (\boldsymbol{\mu}_j,\Sigma_j) & \stackrel{ind.}{\sim} & N(\boldsymbol{\mu}_j,\Sigma_j), \,\,\, j=1,2, \,\,\, \ell=1,\ldots,L \\
\boldsymbol{\gamma}_{\ell} & \stackrel{i.i.d.}{\sim} & N(\boldsymbol{\gamma}_0,\Gamma_0),
\,\,\, \ell = 1,\ldots,L-1 \\
(\boldsymbol{\mu}_j,\Sigma_j) & \stackrel{ind.}{\sim} &
N(\boldsymbol{\mu}_j \mid \boldsymbol{\mu}_{0j},\Sigma_j/\kappa_{0j}) \,
IW(\Sigma_j \mid \nu_{0j},\Lambda_{0j}^{-1}),\,\,\, j=1,2\\
\sigma_j^2 & \stackrel{i.i.d.}{\sim} & IG(a_{\sigma},b_{\sigma}),\,\,\, j = 1,2
\end{array}
\end{equation*}
where $p_{d\ell}=$ $\varphi(\mathbf{x}_d^T \boldsymbol{\gamma}_{\ell}) 
\prod_{h=1}^{\ell-1} (1-\varphi(\mathbf{x}_d^T\boldsymbol{\gamma}_{h}))$, 
for $\ell = 1,\ldots,L-1$, and $p_{dL}=$
$\prod_{\ell=1}^{L-1} (1-\varphi(\mathbf{x}_d^T\boldsymbol{\gamma}_{\ell}))$.

The hierarchical model formulation reminds us the Pólya-Gamma data augmentation approach  \citep{Polson2013}, which is the key to conditionally conjugate updates for all parameters. For each response $(R_{di},y_{di})$, we introduce two groups of Pólya-Gamma latent variables $\boldsymbol{\xi}_{di}=(\xi_{di1},\ldots,\xi_{di,L-1})$ and $\boldsymbol{\zeta}_{di}=(\zeta_{di1},\zeta_{di2})$, where $\xi_{di\ell}\stackrel{i.i.d.}{\sim}PG(1,0)$ and $\zeta_{di1} \sim PG(m_{di},0)$, $\zeta_{di2} \sim PG(m_{di}-R_{di},0)$, independently. Here $PG(b,c)$ denotes the Pólya-Gamma distribution with shape parameter $b$ and tilting parameter $c$. 

We outline the MCMC sampling algorithm for the full augmented model. This process can be achieved entirely with Gibbs updates, by iterating the following steps. For notation simplicity, we let $(\phi\mid -)$ denote the posterior full conditional distribution for parameter $\phi$. 

 \begin{description}
   \item[Step 1: update parameters in the atoms.] In this step, we update two sets of parameters, $\{\boldsymbol{\psi}_{di}:d=1,\ldots,N,i=1,\ldots,n_d\}$, $\{\boldsymbol{\zeta}_{di}:d=1,\ldots,N, i=1,\ldots,n_d\}$ and $\{\boldsymbol{\beta}_{j\ell}:j=1,2,\ell=1,\ldots,L\}$. Denote the set of distinct values of the configuration variables by $\{\mathcal{L}_r^*: r=1,\ldots,n^*\}$. Following \citet{Polson2013}, it can be done by $(\psi_{dij}\mid -)\sim N(\phi_{dij},\tau^2_{dij})$, with $\phi_{di1}=\frac{\mathbf{x}^{\top}\boldsymbol{\beta}_{1\mathcal{L}_{di}}+\sigma_1^2(R_{di}-\frac{m_{di}}{2})}{1+\sigma_1^2\zeta_{di1}}$, $\phi_{di2}=\frac{\mathbf{x}^{\top}\boldsymbol{\beta}_{2\mathcal{L}_{di}}+\sigma_2^2(y_{di}-\frac{m_{do}-R_{di}}{2})}{1+\sigma_2^2\zeta_{di2}}$, $\tau_{dij}^2=\frac{\sigma_j^2}{1+\sigma_j^2\zeta_{dij}}$, $j=1,2$. It is then followed by updating
$(\zeta_{di1}\mid -)\sim PG(m_{di},\psi_{di1})$ and $(\zeta_{di2}\mid -)\sim PG(m_{di}-R_{di},\psi_{di2})$. 
 Finally, we update $\boldsymbol{\beta}_{j\ell}$ by sampling from $(\boldsymbol{\beta}_{j\ell}\mid -)\sim N(\tilde{\boldsymbol{\mu}}_{j\ell},\tilde{\Sigma}_{j\ell})$, where
   \begin{equation*}
   	\begin{array}{rcl}
   		& \bullet\,\,\,\text{if}\,\, \ell\notin\{\mathcal{L}_r^*: r=1,\ldots,n^*\}: & \tilde{\boldsymbol{\mu}}_{j\ell}=\boldsymbol{\mu}_j,\\
   		& & \tilde{\Sigma}_{j\ell}=\Sigma_j;\\
   		& \bullet\,\,\, \text{if}\,\, \ell\in\{\mathcal{L}_r^*: r=1,\ldots,n^*\}: & 
   		\tilde{\boldsymbol{\mu}}_{j\ell}=\tilde{\Sigma}_{j\ell}(\frac{X_{\ell}^T\boldsymbol{\psi}_{\ell}}{\sigma_j^2}+\Sigma_j^{-1}\boldsymbol{\mu}_j),\\
   		& & \tilde{\Sigma}_{j\ell}=(\frac{X_{\ell}^TX_{\ell}}{\sigma_j^2}+\Sigma_j^{-1})^{-1}.
   	\end{array}
   \end{equation*}
   Here $X_{\ell}$ is the matrix whose column vectors are given by $\{\mathbf{x}_d:\mathcal{L}_{di}=\ell\}$, and $\boldsymbol{\psi}_{\ell}$ is the vector of $\{\psi_{dij}:\mathcal{L}_{di}=\ell\}$. Notice that updating $\{\boldsymbol{\beta}_{j\ell}\}$ can be run in parallel across categories $j=1,\ldots, C-1$. 
   
   \item[Step 2: update parameters in the weights.] Similarly, we update $\{\boldsymbol{\gamma}_{\ell}:\ell=1,\ldots,L-1\}$ and $\{\xi_{i\ell}:i=1,\ldots,n,\ell=1,\ldots,L-1\}$ from $(\boldsymbol{\gamma}_{\ell}\mid -)\sim N(\tilde{\boldsymbol{\gamma}}_{\ell},\tilde{\Gamma}_{\ell})$ and $(\xi_{i\ell}\mid -)\sim PG(1,\mathbf{x}_i^T\boldsymbol{\gamma}_{\ell})$, where $\tilde{\boldsymbol{\gamma}}_{\ell}=\tilde{\Gamma}_{\ell}(X^T_{\ell}\boldsymbol{\iota}_{\ell}+\Gamma^{-1}_0\boldsymbol{\gamma}_0)$ and $\tilde{\Gamma}_{\ell}=(X^T_{\ell}\Xi_{\ell}X_{\ell}+\Gamma_0^{-1})^{-1}$. We denote the diagonal matrix formed by $\{\xi_{i\ell}:\mathcal{L}_i=\ell\}$ as $\Xi_{\ell}$, and the vector of$\{\iota_{i\ell}: \mathcal{L}_i=\ell\}$ as $\boldsymbol{\iota}_{\ell}$. 
   
   \item[Step 3: update configuration variables.]  Update $\mathcal{L}_i$, for $i=1,\ldots,n$ from
   \begin{equation*}
       P(\mathcal{L}_i=\ell\mid -)=\frac{p_{i\ell}\prod_{j=1}^{C-1}Bin(Y_{ij}\mid m_{ij},\varphi(\boldsymbol{x}_i^T\boldsymbol{\beta}_{j\ell}))}{\sum_{\ell=1}^Lp_{i\ell}\prod_{j=1}^{C-1}Bin(Y_{ij}\mid m_{ij},\varphi(\boldsymbol{x}_i^T\boldsymbol{\beta}_{j\ell}))}
   \end{equation*}
   where $\{p_{i\ell}: \ell=1,\ldots, L\}$ are calculated as $p_{i1}=\varphi(\mathbf{x}_i^T\boldsymbol{\gamma}_{1})$, $p_{i\ell}=\varphi(\mathbf{x}_i^T\boldsymbol{\gamma}_{\ell})\prod_{h=1}^{\ell-1}(1-\varphi(\mathbf{x}_i^T\boldsymbol{\gamma}_h))$, $\ell=2,\ldots,L-1$, and  $p_{iL}=\prod_{\ell=1}^{L-1}$ $(1-\varphi(\mathbf{x}_i^T\boldsymbol{\gamma}_{\ell}))$.
   
   \item[Step 4: update overdispersion parameters.] The posterior full conditional distribution of $\sigma_j^2$ is given by $(\sigma^2_j\mid -)\sim IG(a_j^*,b_j^*)$, where
   \begin{equation*}
   	 a_j^*=a_{\sigma}+\frac{n_j^*}{2},\,\,\, b_j^*=b_{\sigma}+\frac{\sum_{d=1}^N\sum_{i=1}^{n_d}(\psi_{dij}-\mathbf{x}_d^{\top}\boldsymbol{\beta}_{j\mathcal{L}_{di}})}{2},
   \end{equation*}
   where $n_1^*=\sum_{d=1}^Nn_d$ and $n_2^*=\sum_{d=1}^N\sum_{i=1}^{n_d}\mathbf{1}(m_{di}-R_{di}>0)$.
   
   \item[Step 5: update hyperparameters.] By conjugacy, updating hyperparameters is standard. We update $\{\boldsymbol{\mu}_j\}$ and $\{\Sigma_j\}$ by $(\boldsymbol{\mu}_j\mid -)\sim N(\boldsymbol{\mu}^*_j,\Sigma_j/\kappa^*_j)$ and $(\Sigma_j\mid -)\sim IW(\nu^*_j,(\Lambda^*_j)^{-1})$, with the parameters given by
   \begin{equation*}
   \begin{split}
    &\boldsymbol{\mu}^*_j=\frac{\kappa_{0j}}{\kappa_{0j}+n^*}\boldsymbol{\mu}_{0j}+\frac{n^*}{\kappa_{0j}+n^*}\bar{\boldsymbol{\beta}}_j,\,\,\,\kappa^*_j=n^*+\kappa_{0j},\,\,\, \nu^*_j=n^*+\nu_{0j}\\
    & \bar{\boldsymbol{\beta}}_j=\frac{1}{n^*}\sum_{r=1}^{n^*}\boldsymbol{\beta}_{j\mathcal{L}_r^*}, \,\,\, S_j=\sum_{r=1}^{n^*}(\boldsymbol{\beta}_{j\mathcal{L}_r^*}-\bar{\boldsymbol{\beta}}_j)(\boldsymbol{\beta}_{j\mathcal{L}_r^*}-\bar{\boldsymbol{\beta}}_j)^T, \\
    &\Lambda^*_j=\Lambda_{0j}+S_j+\frac{n^*\kappa_{0j}}{n^*+\kappa_{0j}}(\bar{\boldsymbol{\beta}}_j-\boldsymbol{\mu}_{0j})(\bar{\boldsymbol{\beta}}_j-\boldsymbol{\mu}_{0j})^T.   
   \end{split}
   \end{equation*}
\end{description}

The posterior sampling algorithm for the ``CW-LNB'' model is adapt from this sampling scheme, with \textbf{Step 2} and \textbf{Step 3} been replaced by \textbf{Step $2^*$} and \textbf{Step $3^*$}, while keeping the other steps unchanged.

\begin{description}
\item[Step $2^*$: update parameters in the weights.] The parameters to be updated in this step involve $\{\omega_{\ell}:\ell=1,\ldots,L-1\}$ and $\alpha$. From \citet{IshwaranJames2001}, it can be done by sample $V_{\ell}^*\stackrel{ind.}{\sim}Beta(1+M_{\ell},\alpha+\sum_{h=\ell+1}^LM_h)$ for $\ell=1,\ldots,L-1$. Then let $\omega_1=V_1^*$, $\omega_{\ell}=V_{\ell}^*\prod_{h=1}^{\ell-1}(1-V_h^*)$, $\ell=2,\ldots,L-1$ and $\omega_L=1-\sum_{\ell=1}^{L-1}\omega_{\ell}$. In addition, a new sample of $\alpha$ is obtained from $(\alpha\mid -)\sim Gamma(a_{\alpha}+L-1,b_{\alpha}-\sum_{\ell=1}^{L-1}\log(1-V_{\ell}^*))$.

\item[Step $3^*$: update configuration variables.] Update $\mathcal{L}_i$, $i=1,\ldots,n$, from
\begin{equation*}
   P(\mathcal{L}_i=\ell\mid -)=\frac{\omega_{\ell}\prod_{j=1}^{C-1}Bin(Y_{ij}\mid m_{ij},\varphi(\boldsymbol{x}_i^T\boldsymbol{\beta}_{j\ell}))}{\sum_{\ell=1}^L\omega_{\ell}\prod_{j=1}^{C-1}Bin(Y_{ij}\mid m_{ij},\varphi(\boldsymbol{x}_i^T\boldsymbol{\beta}_{j\ell}))}. 
\end{equation*}
\end{description}

\section{Proofs}
\label{sec:proof}

\subsection{Proof of Proposition \ref{prop:lnbmargapprox}}

\begin{proof}
	By definition, we have
	\begin{equation*}
		\text{Corr}(\tilde{Y}_q,\tilde{Y}_{q^{\prime}}\mid \theta,\sigma^2)\,=\,\frac{\text{E}(\tilde{Y}_q\tilde{Y}_{q^{\prime}}\mid \theta,\sigma^2)-\text{E}(\tilde{Y}_q\mid \theta,\sigma^2)\text{E}(\tilde{Y}_{q^{\prime}}\mid \theta,\sigma^2)}{\sqrt{\text{Var}(\tilde{Y}_q\mid \theta,\sigma^2)\text{Var}(\tilde{Y}_{q^{\prime}}\mid \theta,\sigma^2)}}.
	\end{equation*}
	Using law of total expectation/variance,
	\begin{equation*}
		\begin{array}{rcl}
			\text{E}(\tilde{Y}_q\tilde{Y}_{q^{\prime}}\mid \theta,\sigma^2) & = &
			\text{E}(\varphi^2(\psi)\mid\theta,\sigma^2);\\
			\text{E}(\tilde{Y}_q\mid \theta,\sigma^2)\,=\,\text{E}(\tilde{Y}_{q^{\prime}}\mid \theta,\sigma^2) & = & \text{E}(\varphi(\psi)\mid\theta,\sigma^2);\\
			\text{Var}(\tilde{Y}_q\mid \theta,\sigma^2)\,=\,\text{Var}(\tilde{Y}_{q^{\prime}}\mid \theta,\sigma^2) & = & \text{E}(\varphi(\psi)\mid\theta,\sigma^2)-\{\text{E}(\varphi(\psi)\mid\theta,\sigma^2)\}^2,
		\end{array}
	\end{equation*}
	where the expectation is taken with respect to $\psi\sim N(\theta,\sigma^2)$. 
	
	Write $\psi=\theta+\zeta$, where $\zeta\sim N(0,\sigma^2)$. By Taylor expansion around the mean, 
	\begin{equation*}
		\varphi(\psi)\approx \varphi(\theta)+\zeta\varphi^{\prime}(\theta)+\frac{\zeta^2}{2}\varphi^{\prime\prime}(\theta).
	\end{equation*}
	Then taking expectation yields $\text{E}(\varphi(\psi)\mid\theta,\sigma^2)\approx$ 
	$\varphi(\theta)+\frac{\sigma^2}{2}\varphi^{\prime\prime}(\theta)$. Using the same technique, 
	\begin{equation*}
		\varphi^2(\psi)\approx \varphi^2(\theta)+2\zeta\varphi(\theta)\varphi^{\prime}(\theta)+\zeta^2\{(\varphi^{\prime}(\theta))^2+\varphi(\theta)\varphi^{\prime\prime}(\theta)\}.
	\end{equation*}
	Taking expectation yields $\text{E}(\tilde{Y}_q\tilde{Y}_{q^{\prime}}\mid \theta,\sigma^2)\approx$ $\varphi^2(\theta)+\sigma^2\{(\varphi^{\prime}(\theta))^2+\varphi(\theta)\varphi^{\prime\prime}(\theta)\}$ .
	
	We further notice that $\varphi^{\prime}(\theta)=$ $\varphi(\theta)(1-\varphi(\theta))$, and $\varphi^{\prime}(\theta)=$ $\varphi(\theta)(1-\varphi(\theta))(1-2\varphi(\theta))$. The final results emerge after simple algebra. 
\end{proof}

\subsection{Proof of Proposition \ref{prop:equmgforimodel}}
\label{subsec:proofequmgforimodel}

\begin{proof}
Starting with the moment generating function for $\sum_{q=1}^m\tilde{R}_q$ and $\sum_{l=1}^{m-\sum_q\tilde{R}_q}\tilde{y}_l$ under model $\tilde{\mathcal{M}}$, we have 
{\allowdisplaybreaks
\begin{align*}
       E&_{\tilde{\mathcal{M}}}(e^{t_1\sum \tilde{R}_q+t_2\sum \tilde{y}_l}\mid
        m,G_{\mathbf{x}})\\
       &=\sum_{\ell=1}^{\infty}\omega_{\ell}(\mathbf{x})\,
       \biggl\{\prod_{q=1}^m \sum_{\tilde{R}_q=0,1}e^{t_1\tilde{R}_q}Bern(\tilde{R}
       _q\mid\varphi(\theta_{1\ell}(\mathbf{x})))\biggr\}\biggl\{\prod_{l=1}^{m-\sum_q\tilde{R}_q}\sum_{\tilde{y}_l=0,1}e^{t_2\tilde{y}_l}Bern(\tilde{y}_l\mid \varphi({\theta}_{2\ell}(\mathbf{x})))\biggr\}\\
       &=\sum_{\ell=1}^{\infty}\omega_{\ell}(\mathbf{x})\,\biggl\{\prod_{q=1}^m \sum_{\tilde{R}_q=0,1}e^{t_1\tilde{R}_q}Bern(\tilde{R}_q\mid\varphi(\theta_{1\ell}(\mathbf{x})))\biggr\}\biggl\{1+\varphi(\theta_{2\ell}(\mathbf{x}))(e^{t_2}-1)\biggr\}^{m-\sum\tilde{R}_q}\\
       &=\sum_{\ell=1}^{\infty}\omega_{\ell}(\mathbf{x})\,\biggl[\,\prod_{q=1}^m \sum_{\tilde{R}_q=0,1}\biggl\{e^{t_1\tilde{R}_q}Bern(\tilde{R}_q\mid\varphi(\theta_{1\ell}(\mathbf{x})))\{1+\varphi(\theta_{2\ell}(\mathbf{x}))(e^{t_2}-1)\}^{1-\tilde{R}_q}\biggr\}\,\biggr]\\
       &=\sum_{\ell=1}^{\infty}\omega_{\ell}(\mathbf{x})\,\biggl[(1-\varphi(\theta_{1\ell}(\mathbf{x})))\{1+\varphi(\theta_{2\ell}(\mathbf{x}))(e^{t_2}-1)\}
       +e^{t_1}\varphi(\theta_{1\ell}(\mathbf{x}))\biggr]^m\\
       &=\sum_{\ell=1}^{\infty}\omega_{\ell}(\mathbf{x})\sum_{R=0}^m \binom{m}{R}\biggl\{e^{t_1}\varphi(\theta_{1\ell}(\mathbf{x}))\biggr\}^R
       \biggl[(1-\varphi(\theta_{1\ell}(\mathbf{x})))
       \{1+\varphi(\theta_{2\ell}(\mathbf{x}))(e^{t_2}-1)\}\biggr]^{m-R}\\
       &=\sum_{\ell=1}^{\infty}\omega_{\ell}(\mathbf{x})\,\biggl\{\sum_{R=0}^m e^{t_1R}Bin(R\mid m,\varphi(\theta_{1\ell}(\mathbf{x})))\biggr\}\biggl[1+\varphi(\theta_{2\ell}(\mathbf{x}))(e^{t_2}-1)\biggr]^{m-R}\\
       &=\sum_{\ell=1}^{\infty}\omega_{\ell}(\mathbf{x})\,\biggl\{\sum_{R=0}^m e^{t_1R}Bin(R\mid m,\varphi(\theta_{1\ell}(\mathbf{x})))\biggr\}\biggl\{\sum_{y=0}^{m-R} e^{t_2y}Bin(y\mid m-R,\varphi(\theta_{2\ell}(\mathbf{x})))\biggr\}\\
       &=E_{\mathcal{M}}(e^{t_1R+t_2y}\mid m,G_{\mathbf{x}}),
\end{align*}}
which completes the argument for the proof.
\end{proof}

\subsection{Proof of Proposition \ref{prop:overdequalmgf}}
\begin{proof}

We first show the equality holds on the parametric backbones of the corresponding nonparametric mixture models. Let $\mathcal{M}^*$ and $\tilde{\mathcal{M}}^*$ denote the kernel of the mixture model $\mathcal{M}$ and $\tilde{\mathcal{M}}$, respectively. Following the strategy in proving Proposition \ref{prop:equmgforimodel}, we start from the MGF for $\sum_{q=1}^m\tilde{R}_q$ and $\sum_{l=1}^{m-\sum_q\tilde{R}_q}\tilde{y}_l$,
	{\allowdisplaybreaks
\begin{align*}
       E&_{\tilde{\mathcal{M}}^*}(e^{t_1\sum \tilde{R}_q+t_2\sum \tilde{y}_l}\mid
        m,\theta_1(\mathbf{x}),\theta_2(\mathbf{x}),\boldsymbol{\sigma^2})\\
       &=\biggl\{\int \prod_{q=1}^m \sum_{\tilde{R}_q=0,1}e^{t_1\tilde{R}_q}Bern(\tilde{R}
       _q\mid\varphi(\psi_1))N(\psi_1\mid \theta_1(\mathbf{x}),\sigma_1^2)d\psi_1\biggr\}\\
       &\quad\times\biggl\{\int \prod_{l=1}^{m-\sum_q\tilde{R}_q}\sum_{\tilde{y}_l=0,1}e^{t_2\tilde{y}_l}Bern(\tilde{y}_l\mid \varphi(\psi_2))N(\psi_2\mid\theta_2(\mathbf{x}),\sigma_2^2)d\psi_2\biggr\}\\
       &=\int\int\biggl\{\prod_{q=1}^m \sum_{\tilde{R}_q=0,1}e^{t_1\tilde{R}_q}Bern(\tilde{R}
       _q\mid\varphi(\psi_1))\biggr\}\biggl\{\prod_{l=1}^{m-\sum_q\tilde{R}_q}\sum_{\tilde{y}_l=0,1}e^{t_2\tilde{y}_l}Bern(\tilde{y}_l\mid \varphi(\psi_2))\biggr\}\\
       &\quad \times N(\psi_1\mid \theta_1(\mathbf{x}),\sigma_1^2)N(\psi_2\mid\theta_2(\mathbf{x}),\sigma_2^2)d\psi_1d\psi_2\\
       &=\int\int \biggl\{ \sum_{R=0}^m e^{t_1R}Bin(R\mid m,\varphi(\psi_1))\biggr\}
       \biggl\{ \sum_{y=0}^{m-R} e^{t_2y}Bin(y\mid m-R,\varphi(\psi_2))\biggr\}\\
       &\quad \times N(\psi_1\mid \theta_1(\mathbf{x}),\sigma_1^2)N(\psi_2\mid\theta_2(\mathbf{x}),\sigma_2^2)d\psi_1d\psi_2\\
       &=\int \biggl\{ \sum_{R=0}^m e^{t_1R}Bin(R\mid m,\varphi(\psi_1))\biggr\}N(\psi_1\mid \theta_1(\mathbf{x}),\sigma_1^2)d\psi_1\\
       &\quad\times \int \biggl\{ \sum_{y=0}^{m-R} e^{t_2y}Bin(y\mid m-R,\varphi(\psi_2))\biggr\}N(\psi_2\mid\theta_2(\mathbf{x}),\sigma_2^2)
       d\psi_2\\
       &=E_{\mathcal{M}^*}(e^{t_1R+t_2y}\mid m,\theta_1(\mathbf{x}),\theta_2(\mathbf{x}),\boldsymbol{\sigma^2}).
\end{align*}}

Turning to the nonparametric mixture model, applying the equality for parametric model on every mixing component, we have

{\allowdisplaybreaks
\begin{align*}
	E&_{\tilde{\mathcal{M}}}(e^{t_1\sum \tilde{R}_q+t_2\sum \tilde{y}_l}\mid
        m,G_{\mathbf{x}},\boldsymbol{\sigma^2})\\
        &=\sum_{\ell=1}^{\infty}\omega_{\ell}(\mathbf{x})\biggl\{\int \prod_{q=1}^m \sum_{\tilde{R}_q=0,1}e^{t_1\tilde{R}_q}Bern(\tilde{R}
       _q\mid\varphi(\psi_1))N(\psi_1\mid \theta_{1\ell}(\mathbf{x}),\sigma_1^2)d\psi_1\biggr\}\\
       &\quad \times\biggl\{\int \prod_{l=1}^{m-\sum_q\tilde{R}_q}\sum_{\tilde{y}_l=0,1}
       e^{t_2\tilde{y}_l}Bern(\tilde{y}_l\mid 
       \varphi(\psi_2))N(\psi_2\mid\theta_{2\ell}(\mathbf{x}),\sigma_2^2)d\psi_2\biggr\}\\
       &=\sum_{\ell=1}^{\infty}\omega_{\ell}(\mathbf{x})\biggl[\int \biggl\{ \sum_{R=0}^m e^{t_1R}Bin(R\mid m,\varphi(\psi_1))\biggr\}
       N(\psi_1\mid \theta_{1\ell}(\mathbf{x}),\sigma_1^2)d\psi_1\biggr]\\
       &\quad \times \biggl[\int \biggl\{ \sum_{y=0}^{m-R} e^{t_2y}Bin(y\mid m-R,\varphi(\psi_2))\biggr\}
       N(\psi_2\mid\theta_{2\ell}(\mathbf{x}),\sigma_2^2)d\psi_2\biggr]\\
       &=E_{\mathcal{M}}(e^{t_1R+t_2y}\mid m,G_{\mathbf{x}},\boldsymbol{\sigma^2}).
\end{align*}
}
Therefore, we obtaine the desired equation. 
\end{proof}

\section{Additional Results for Data Examples}
\label{sec:moredataresults}

\subsection{Second Synthetic Data Example}

\begin{figure}[t!]
\centering
\includegraphics[width=14.8cm,height=3.7cm]{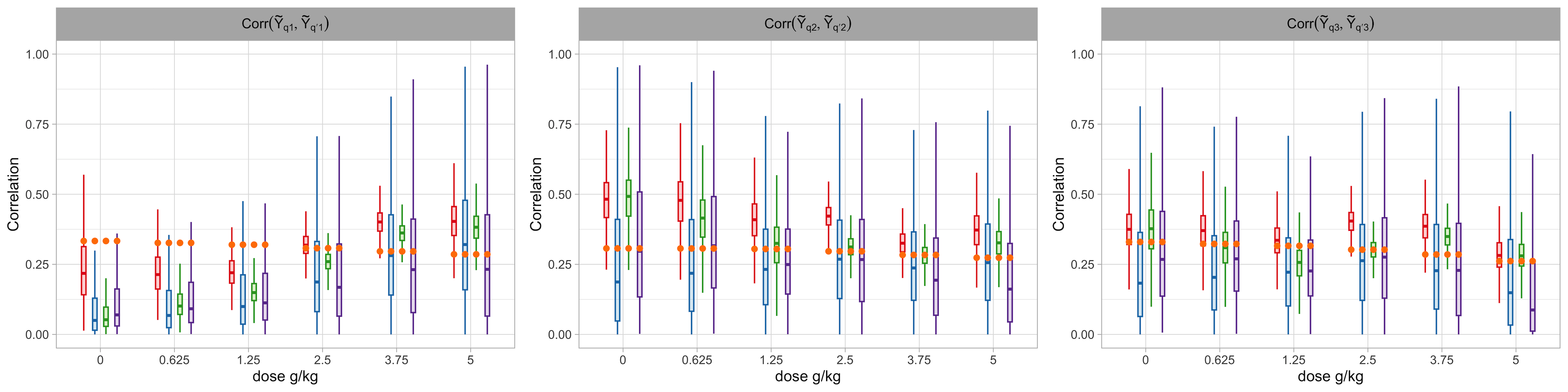}  
\caption{\small Second simulation example. Box plot of the intracluster correlation posterior distributions at the observed toxin levels. In each panel, estimates under the ``CW-Bin'', ``CW-LNB'', ``Gen-Bin'' and ``Gen-LNB'' model are shown in red, blue, green, and purple, respectively. The orange dot marks the truth.}
\label{fig:sim3corr}
\end{figure}

To further investigate how the nonparametric mixture models behave in capturing the overdispersion of the data, we plot in Figure \ref{fig:sim3corr} posterior samples of the intracluster correlation. To facilitate comparison, we calculate the true intracluster correlations from the data generating process, and add them to the plot. In general, the models with overdispersed kernel perform better in capturing the truth. Without the help from overdispersed kernel, the ``CW-Bin'' model and the ``Gen-Bin'' model rely on the mixture structure to account for the extra dispersion, and are less effective in capturing the variability.  

\begin{figure}[t!]
\centering
\includegraphics[width=14.8cm,height=3.7cm]{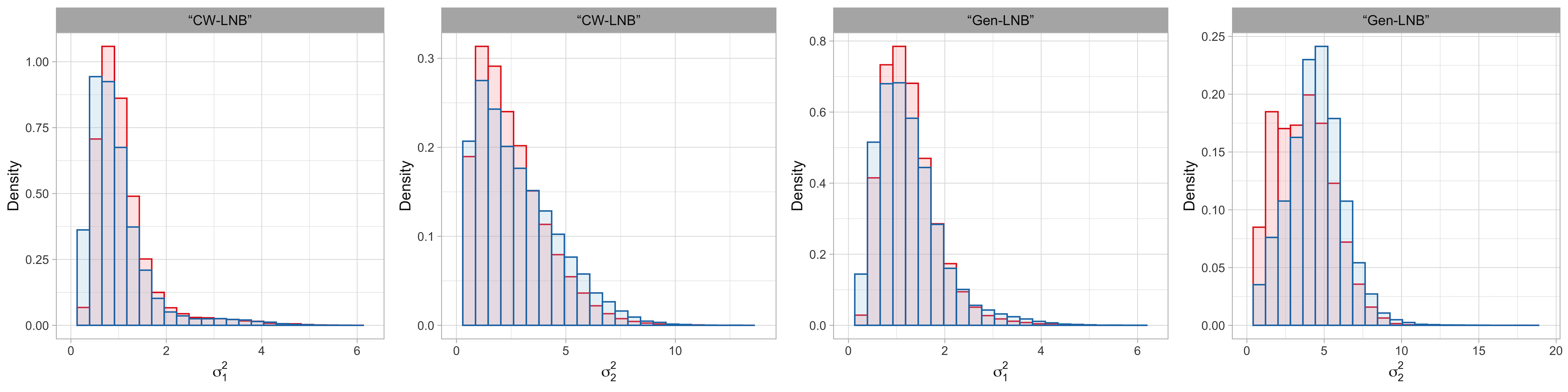}  
\caption{\small Second simulation example. Posterior distributions of the overdispersion parameters $\boldsymbol{\sigma^2}$ under the ``CW-LNB'' and the ``Gen-LNB'' model with prior $IG(3,8/3)$ (in red) and $IG(2,4/3)$ (in blue).}
\label{fig:sim3senana}
\end{figure}

Finally, we conduct a sensitivity analysis regarding the prior of $\boldsymbol{\sigma^2}$ under the ``CW-LNB'' and ``Gen-LNB'' models. The results discussed above correspond to prior $\sigma_j^2\sim IG(3,8/3)$, for $j=1,2$, leading to an extra 33\% variance on average a priori. We consider a more diffuse prior, namely $IG(2,4/3)$, which induces the same level of extra variance on average. The posterior distributions of $\boldsymbol{\sigma^2}$ are shown in Figure \ref{fig:sim3senana}. These distributions are comparable (and different from the prior distribution), suggesting effective learning of overdispersion from the data.

\subsection{Real Data Example}

As an illustration of assessing convergence of the MCMC algorithm, we provide some results with regard to  the ``Gen-LNB'' model for the real data example. For this example, a total of 5000 MCMC samples were taken from a chain of length 30000. The first 20000 were discarded as burn in, and the remaining draws were thinned to reduce autocorrelation. 

\begin{figure}[t!]
\centering
\includegraphics[width=14.8cm,height=3.7cm]{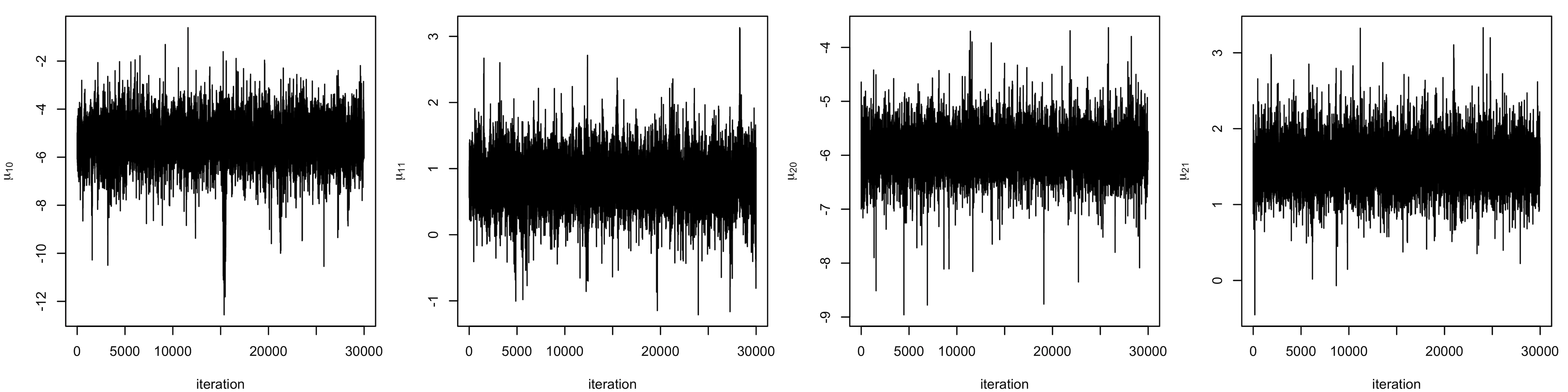}  
\caption{\small MCMC diagnostics for the real data example. Trace plots of 
the complete chain of posterior samples for the elements of $\boldsymbol{\mu}_j$, $j=1,2$.}
\label{fig:MCMCdiagmu}
\end{figure}

\begin{figure}[t!]
\centering
\includegraphics[width=14.8cm,height=7.4cm]{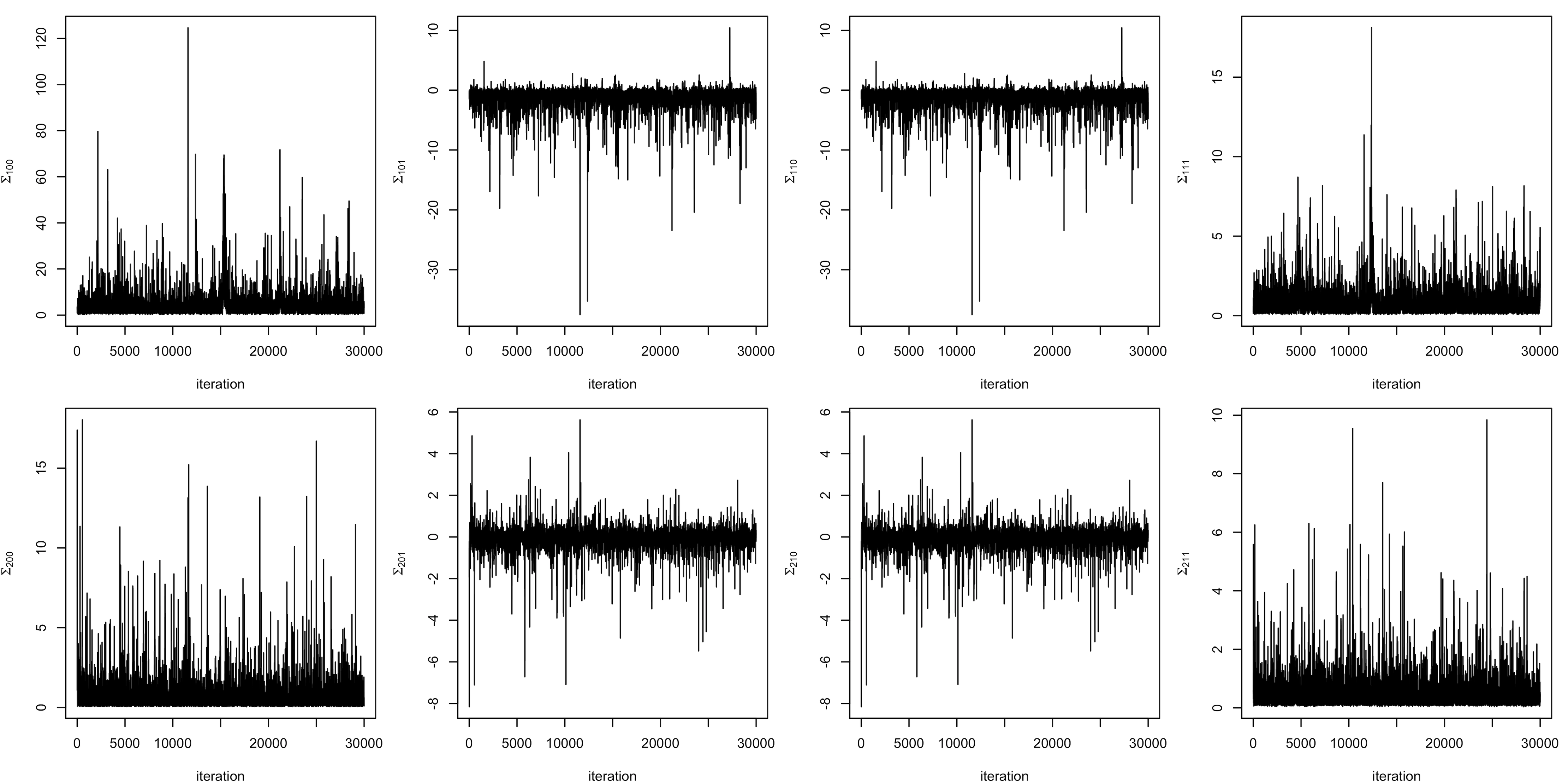}  
\caption{\small MCMC diagnostics for the real data example. Trace plots of 
the  complete chain of posterior samples for the elements of $\Sigma_j$, $j=1,2$.}
\label{fig:MCMCdiagSigma}
\end{figure}

Figure \ref{fig:MCMCdiagmu} and Figure \ref{fig:MCMCdiagSigma} show the trace plots of the 
complete chain of posterior samples for the elements of $\boldsymbol{\mu}_j$ and $\Sigma_j$, $j=1,2$, 
respectively. The plots suggest that convergence is achieved quickly for these parameters, though 20000 appears to be a conservatively safe burn-in interval.

\begin{figure}[t!]
\centering
\includegraphics[width=14.8cm,height=3.7cm]{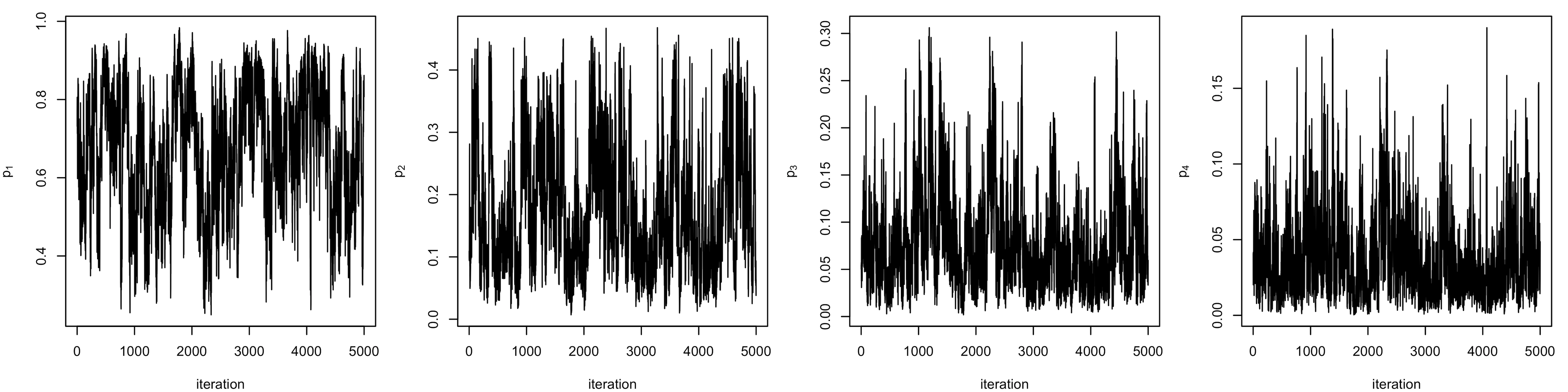}  
\caption{\small MCMC diagnostics for the real data example. Trace plots of the 5000 remaining posterior samples for the four largest elements of the mixing weight vector.}
\label{fig:MCMCdiagweight}
\end{figure}

The remaining trace plots focus on the 5000 MCMC samples collected after burn in and 
thinning. Figure \ref{fig:MCMCdiagweight} shows the trace plots for the largest 4 elements 
of the mixing weight vector $\{p_{d\ell}:\ell=1,\ldots,L\}$, for a randomly selected dose level $x_d$, where $p_{d1}\,=\,\varphi(\mathbf{x}_d^{\top}\boldsymbol{\gamma}_1)$, $p_{d\ell}\,=\,\varphi(\mathbf{x}_d^{\top}\boldsymbol{\gamma}_{\ell}) 
\prod_{h=1}^{\ell-1} (1-\varphi(\mathbf{x}_d^{\top}\boldsymbol{\gamma}_h))$, $\ell=2,\ldots,L-1$, and $p_{dL}=1-\sum_{\ell=1}^{L-1}p_{d\ell}$.  
Specifically, at each MCMC iteration the largest element of the vector of mixture weights 
is found, and this corresponds to $p_1$ in the plot. Similarly, 
the second largest element at each iteration is found, and this corresponds to $p_2$ in the plot,
with $p_3$ and $p_4$ obtained in the same fashion.

\begin{figure}[t!]
\centering
\includegraphics[width=14.8cm,height=3.7cm]{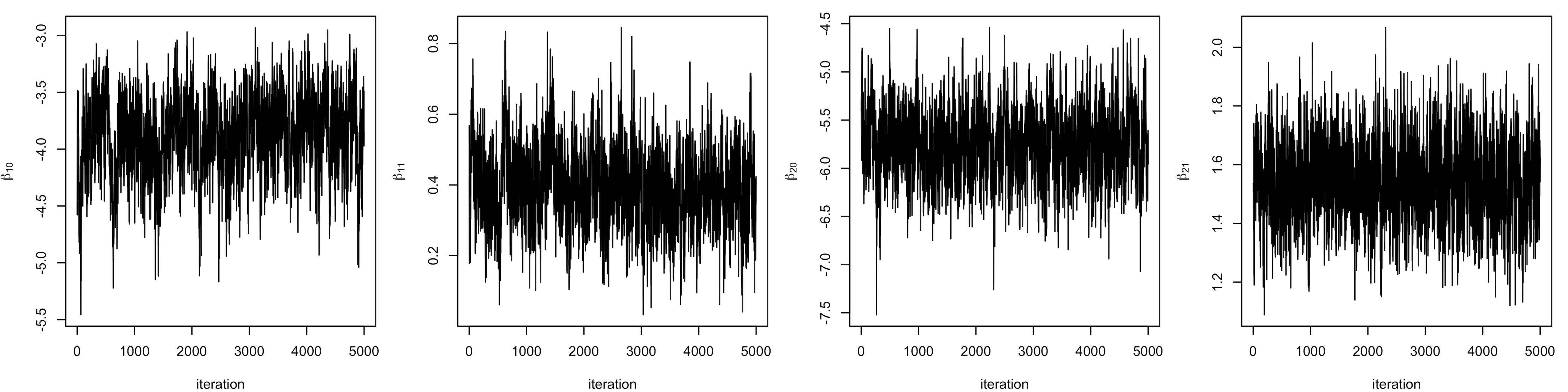}  
\caption{\small MCMC diagnostics for the real data example. Trace plots of the 5000 
remaining posterior samples for the label-invariant functions of $\boldsymbol{\beta}_{j\ell}$.}
\label{fig:MCMCdiagbeta}
\end{figure}

Because of the label switching problem in mixture models, we turn to label-invariant 
functions of parameters $\boldsymbol{\beta}_{j\ell}$ to diagnose convergence and mixing for these 
parameters. We use $\sum_{d=1}^N\sum_{i=1}^{n_d}\mathbf{\beta}_{j\mathcal{L}_{di}}/n$, where 
$\mathcal{L}_{di}$ represents the configuration variable corresponding to subject $i$ in dose group $d$, and $n=\sum_{d=1}^Nn_d$. This quantity 
can be interpreted as the average of the coefficient vectors to which the observations are 
assigned at a particular MCMC iteration, or a weighted average of the $\boldsymbol{\beta}_{j\ell}$
parameters. It is commonly used to assess convergence in mixture models 
\citep[see, e.g.,][]{Antoniano2016}. Figure \ref{fig:MCMCdiagbeta} shows the trace plots for 
the components of the label-invariant functions of $\boldsymbol{\beta}_{j\ell}$, which also 
suggests convergence of the Markov chain. 

\begin{figure}[t!]
\centering
\includegraphics[width=14.8cm,height=3.7cm]{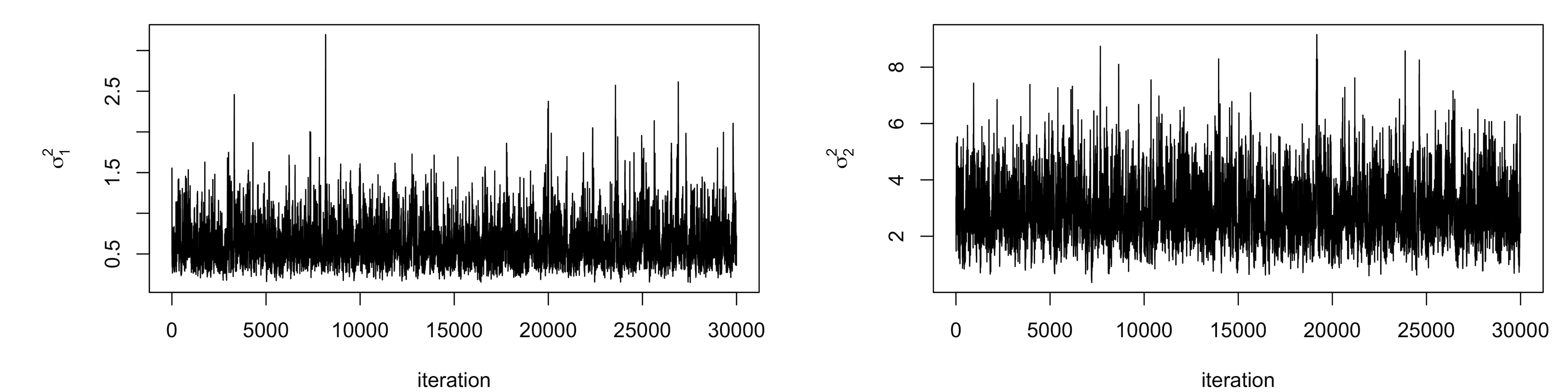}  
\caption{\small MCMC diagnostics for the real data example. Trace plots of the 5000 
remaining posterior samples for the overdispersed parameter $\sigma_1^2$ and $\sigma_2^2$.}
\label{fig:MCMCdiagsigmasq}
\end{figure}

\begin{figure}[t!]
\centering
\includegraphics[width=14.8cm,height=3.7cm]{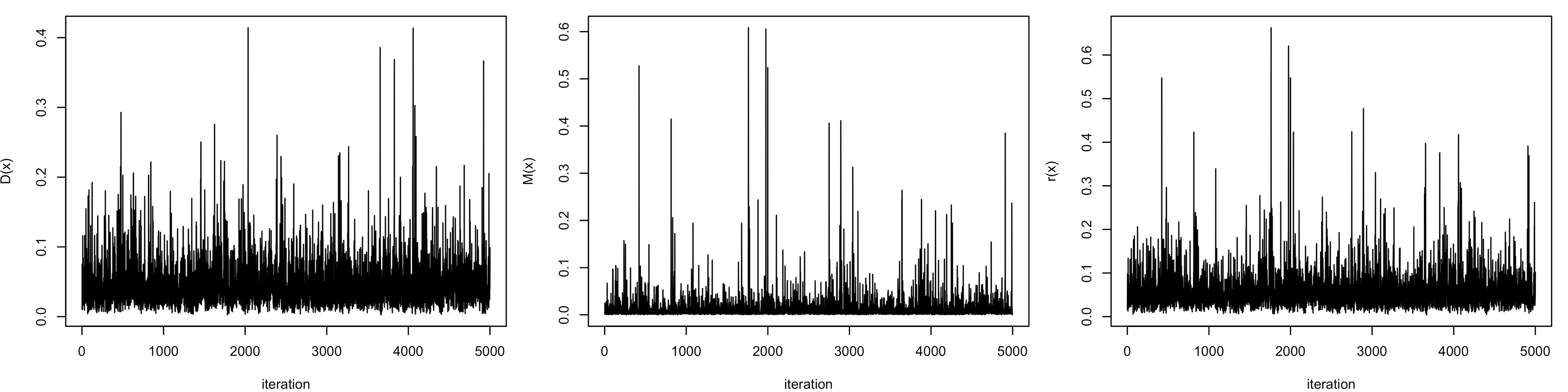}  
\caption{\small MCMC diagnostics for the real data example. Trace plots of the 5000 
remaining posterior samples for the dose-response probabilities $D(x)$, $M(x)$ and $r(x)$, corresponding to a randomly selected $x_d$.}
\label{fig:MCMCdiagprob}
\end{figure}

Moreover, Figure \ref{fig:MCMCdiagsigmasq} shows the trace plots of the posterior samples of overdispersed parameters, and Figure \ref{fig:MCMCdiagprob} displays the trace plots for the estimated dose-response 
probabilities $D(x)$, $M(x)$ and $r(x)$, for a randomly chosen dose level ${x}_d$. Again, these plots suggest no illness regrarding the MCMC convergence. 

